\documentclass[11pt]{article}
\usepackage{amsmath}
\usepackage{amssymb}

\input{tcilatex}
\begin{document}

\title{Fair Congested Assignment Problem\thanks{\textit{Acknowledgments}: We
thank Felix Fischer, Ioannis Caragiannis, Igal Milchtaich and Sinan Ertemel
for critical comments and references. }}
\author{Anna Bogomolnaia\thanks{%
Adam Smith Business School, University of Glasgow and Centre d'Economie de
la Sorbonne, CNRS, Paris} \ and Herv\'{e} Moulin\thanks{%
Adam Smith Business School, University of Glasgow}}
\date{February 2024}
\maketitle

\begin{abstract}
We must assign $n$ agents to $m$ posts subject to negative congestion: what
assignment is fair and efficient?

If congestion is anonymous (each agent adds one unit) it is always possible
to assign each agent to one of their top $n$ out of the $n\times m$ feasible
allocations. This ordinal interpretation of Ex Ante fairness can be adjusted
if congestion is weighted (agent-specific).

An assignment is \textit{Competitive }if I don't want to move to an empty
post, or to an occupied one \textit{at its current congestion level}. If it
exists the competitive assignment is essentially unique, efficient and Ex
Ante fair.

Among agents endowed with cardinal vNM utilities we can randomise the
selection of our assignment. Under anonymous congestion every problem has a
unique competitive congestion profile, implemented by a mixture of
deterministic assignments rounding up or down the competitive congestion,
and approximately Ex Ante fair, efficient and welfare equivalent. Some of
these properties are lost under weighted congestion.


\end{abstract}

\section{introduction}

Congestion affects the allocation of any commodity consumed under partial
rivalry: traffic on roads and the internet, public utilities and private
services, and so on: for just over a century it is a staple of microeconomic
analysis. Pigou's work on the taxation of externalities(\cite{Pig}) inspired
the search for the welfare-optimal taxation of congestion, for instance peak
load pricing of utilities (\cite{Ram}, \cite{Boi}),Vickrey's optimal toll
road (\cite{Vic}) and the far reaching Vickrey-Clarke-Groves pricing
mechanisms (\cite{GL}).

A different and equally outstanding operations research and game theory
literature starts with Wardrop's transportation model (\cite{CS}):
congestion is analysed in the decentralised \textit{free mobility}\ regime:
each agent chooses her service (what route to use, which store to visit,
etc..) and the congestion results from these non cooperative interactions
(typically without any transfer of money). Such games include the seminal
instance of potential games (\cite{Ros}, \cite{MS}) and the first definition
of the price of anarchy\footnote{%
Measuring the welfare loss in the worst Nash equilibrium of the game; see e.
g. \textbf{\cite{Fra}, }\cite{RT}.} (\cite{KP}, \cite{Rou}).

Our model uses the same physical description of assignments as a famiiar
congestion game (e. g. \cite{Mil}): there is a set of heterogenous items
subject to congestion that we call "posts"; each agent must be assigned to a
post, and each post can take any amount of congestion, from remaining empty
to hosting everyone. The only restriction on agents' preferences is that at
any given post they decrease strictly in the amount of congestion.
Critically cash transfers (taxes) are ruled out: agents cannot be
compensated to accept an inferior post, or pay to be assigned at a popular
one. Our viewpoint is purely normative, non strategic: which of the
efficient assignments should we call fair?

Formal research on this aspect of congestion is scarce and quite recent (see
details in subsection 1.2). Yet moneyless assignment problems where
congestion is an important consideration are easy to find: the allocation of
jobs to busy heterogenous servers, of workers to shared office spaces,
patients to hospitals, students to crowded classes, students to crowded
schools, or messages to routes in a centralised communication network. In
those examples the decentralised free mobility "choose your own post"
approach is clearly impractical (and possibly inefficient). It is also
unpalatable because it gives an unfair advantage to agents who happen to
play the game better: see the discussion of the perverse strategic
implications of the Boston school assignment mechanism (\cite{ES}).

Matching and assignment models involving \textquotedblleft many agents to
one position\textquotedblright\ routinely incorporate \textit{hard}
congestion constraints in the form of upper or lower bounds on the filling
of each post: maximal capacity of a class (\cite{Bud}) or a school (\cite{AS}%
), minimal or maximal quotas for some subsets of students (\cite{BFIM}, \cite%
{HYY}, \cite{MT}, \cite{KK}\textbf{) }etc.. But when these constraints do
not bite or are altogether absent, \textit{soft} congestion still impacts
agents' welfare and choices. Parents will accept more crowded classes if the
school's academic context is better, or vice versa: the recent paper \cite%
{PTZ} on congested school choice (on which more in the next section) offers
clear evidence of this point.

Our definition of fairness, in line with seven decades of microeconomic
literature on the division of private commodities (e. g. \cite{HM}), is
two-fold. First, Ex Ante fairness focuses on the worst case welfare level
that can be guaranteed to each agent, based on their own trade-offs between
posts and congestion but no other information about the preferences of the
other potentially adversarial participants. Second, Ex Post fairness adapts
the familiar Envy-Freeness property (\cite{Fol}, \cite{Var}) under the name
of \textit{Competitiveness}, to account for the fact that the congestion
itself plays the role of a price.

\textit{Our results}. We identify the powerful Ex Ante test of
\textquotedblleft top-fairness\textquotedblright\ relying only on ordinal
preferences; it is always feasible and often met by only a handful of
assignments. In the deterministic model the Competitiveness property is not
always feasible, but if it is it selects a unique, fair and efficient
solution. When we randomise assignments and the agents have cardinal vNM
utilities, there is always a unique \textit{fractional} competitive
congestion profile, implemented by approximately fair and efficient
deterministic assignments.

\subsection{overview of the paper}

The deterministic model with ordinal preferences, relevant until section 5,
is defined in the short section 2. If congestion is anonymous (unweighted)
an agent's allocation is a pair $(a,s_{a})$ where $a$ is the post and $s_{a}$
the number of agents at $a$. If each agent $i$ brings $w_{i}$ units of
congestion to any post an allocation is $(a,w_{S_{a}})$ where $w_{S_{a}}$ is
the total weight of the set $S_{a}$ of the agents assigned to $a$. In both
cases preferences decrease strictly in the congestion at $a$.

In section 3 we have $m$ posts, $n$ agents and congestion is anonymous; this
gives $m\times n$ distinct allocations. An assignment is top-$n$-fair if
every agent $i$'s allocation is one of $i$'s top $n$ allocations (Definition
1). Top-$n$-fairness is always feasible and compatible with efficiency
(Pareto optimality); it cannot be improved: Proposition 1.

In section 4 congestion is weighted and the definition of top-fairness is
more subtle. At the Ex Ante stage agent $i$ knows only their own weight $%
w_{i}$ and the total weight $W$ of all agents, but not their number: so $i$%
's worst case allows for any finite split of $W-w_{i}$. The set of $i$'s
feasible allocations is the union of $m$ intervals $[w_{i},W]$, of size $%
m\times (W-w_{i})$. An assignment is top-$\frac{1}{m}$-fair if each agent $i$
gets one of their top $(W-w_{i})$ allocations (Definition 2).\footnote{%
Under anonymous congestion $n$ is also $\frac{1}{m}$-th of the number of
allocations.} Such assignments always exist (Proposition 2) and this
guarantee cannot be improved.

Competitiveness in the ordinal model is the subject of section 5. Under
anonymous congestion (subsection 5.1) an assignment is competitive if each
agent $i$ weakly prefers their allocation $(a,s_{a})$ to any other
allocation $(x,s_{x})$ at this assignment, including to $(y,1)$ if the post $%
y$ is empty ($s_{y}=0$) (Definition 3). Taking the largest of $s_{x}$ and $1$
as the congestion price of post $x$, the competitive demand for post $a$
contains $s_{a}$ (equals $s_{a}$ if each demand is single-valued). If it
exists the competitive assignment is compelling: fair in the top-$n$-fair
and competitive senses, efficient and (essentially) unique both in terms of
congestion and welfare (Proposition 3).

The definition of competitive assignments when congestion is weighted
(subsection 5.2) is entirely similar (Definition 4); they are also fair,
efficient and unique under minor restrictions on individual preferences
(Definition 5 and Proposition 4).

In both models even with just two posts it is easy to give examples where
competitive assignments do not exist. To turn Competitiveness into an
operational concept of Ex Post fairness, we randomise the selection of the
assignment. We assume that the agents have von Neuman Morgenstern (vNM)
cardinal utilities $u_{i}$ over allocations, just as in Hylland and
Zeckhauser (\cite{HZ}) for the non congested assignment problem.\footnote{%
However we cannot as they did define a competitive price in fiat money
because the worth of a post depends heavily upon its consumption by other
agents.}

Section 6 focuses on the anonymous congestion model. A randomized assignment
generates a \textit{fractional }expected congestion $\sigma _{a}$ at post $a$%
. Agent $i$'s competitive demand at the congestion profile $\sigma =(\sigma
_{x})_{x\in A}$ is the set $D(u_{i},\sigma )$ of posts maximising $%
u_{i}(x,\sigma _{x})$ over $A$. The profile $\sigma $ is competitive if it
obtains as a (finite) convex combination of deterministic assignments where
each agent $i$ is always assigned to a post in $D(u_{i},\sigma )$
(Definition 6). The key fact is that every problem $(A,N,u)$ \textit{has a
unique competitive congestion profile }$\sigma ^{c}$ (Lemma 1). The profile $%
\sigma ^{c}$ is implemented\footnote{%
We use this term in a non strategic sense: it means that $\sigma ^{c}$ is a
convex combination of finitely many determionistic assignments,} by a
lottery over the deterministic assignments described above, and satisfying
two properties: each agent is always assigned to a post in their competitive
demand; the deterministic congestion $s_{a}$ at each post $a$ rounds the
competitive congestion $\sigma _{a}$ up or down to the nearest integer
(Definition 7). Lemma 2 says that the competitive congestion profile can be
so implemented, and Theorem 1 (our second main result together with
Propositions 1,2) explains that each corresponding deterministic assignment
is approximately top-$n$-fair, competitive and efficient, and that they all
have appproximately identical utilities.

We can randomise a weighted congestion problem in exactly the same way, and
postpone this discussion to the Appendix (subsection 8.3) because the
results are weaker. The fractional competitive congestion exists and is
still unique under some qualifications (Definition 8 and lemma 3). But the
deterministic assignments implementing it no longer approximate the
competitive congestion because the relative difference between individual
weights is unbounded.

The remark at the beginning of section 6 explains why the popular
randomisation of the non congested assignment problem by simply extending
each ordinal preference ordering to its (incomplete) stochastic dominance
ordering of random allocations is in fact not useful under congestion.

Concluding comments in section 7 summarise our findings and suggests further
research directions: by adding upper and lower limits to the congestion of
each post; and by relaxing the strictly negative perception of congestion.
The Appendix contains a couple of proofs.

\subsection{related literature}

1) Two independent recent papers introduce congestion in school choice.
Closest to ours is Phan et al. \cite{PTZ} adding the congestion dimension to
the school choice model (\cite{AS}): students' preferences over shools
depend also on their crowding level measured by the per capita resources at
each school. This is formally isomorphic to our model. To manage crowding
the paper adopts a market clearing definition of ex post fairness adapting
envy-freeness like we do. But the school choice viewpoint adds hard capacity
constraints as well as priorities over students for the schools: this
qualifies envy-freeness as \textquotedblleft no justified
envy\textquotedblright\ and their novel concept of Rationing Crowding
Equilibrium (RCE) is more complex than our competitive assignment; also the
proof that a RCE exists is significantly more involved. Remarkably the RCEs
have the semi-lattice structure from which emerges a maximal RCE student
optimal and unique welfarewise, as in the standard school choice model.

Phan et al's model is deterministic and relies on ordinal preferences, so it
cannot take advantage of the convexification offered by randomisation and
vNM cardinal utilities. Still, up to a rounding argument (reminiscent of the
one in our Lemma 2) in the definition of RCEs, Theorem 2 showing that the
RCEs have the same crowding profile is analog to statement $i)$ in our
Proposition 3.

Copland \cite{Cop} also defines a deterministic school choice model where
students have strict ordinal preferences over allocations $(a,s)$; instead
of envy-freeness he imposes what we call the \textit{free mobility}
equilibrium property (on which more below) to balance fairly the congestion
across schools. This property implies our top-$n$-fairness but is
substantially more permissive than envy-freeness. Copland then proposes a
variant of the Deferred Acceptance mechanism (DA with Voluntary Withdrawals)
to take into account the schools' priorities and capacity constraints. So
the formal similarities with our results are fewer than in \cite{PTZ} but
the general viewpoint on congestion is still the same.

The recent follow up by Chen et al. \cite{CGW} show that deciding whether a
deterministic competitive assignment (Definition 3 subsection 5.1) exists
can be done in polynomial time w. r. t. $m$ and $n$, whereas deciding
whether an Envy-Free and top-fair assignment exists is NP-complete: in other
words Competitiveness can easily be applied in large congestion problems.

3) As mentioned in the introduction, an important result for congestion
games apply to precisely our model with preferences decreasing in \textit{%
anonymous} congestion: the non cooperative game where agents choose their
posts independently always has one or more Nash equilibria (\cite{Mil}, \cite%
{KLW}).\footnote{%
This is not true any more if congestion is weighted: \cite{Mil}.} Similar
games play a key role in the hedonic model of coalition formation (\cite{BKS}%
, \cite{BJ}) and local public goods ( e. g. \cite{BLSW1}).

The Free Mobility (FM) equilibria are logically squeezed between our two
fairness concepts: every FM equilibrium is top-fair and every competitive
assignment is a FM equilibrium (the converse statements do not hold). But
multiple FM equilibria with different welfare consequences and no clear
selection is a common situation, therefore the strategic analysis does not
help us find a normatively compelling~single-valued fair assignment. See
section 3 and subsection 5.1 for some examples.

4) The classic combinatorial optimisation problems known as bin packing or
knapsack (\cite{CILM}) discuss like us how to fill bins (posts) with
indivisible balls (agents); but the balls there are just objects and the
concern is about the welfare of the bins (e.g., to respect a capacity
constraint or to minimise the load) while we focus on the welfare of the
balls and treat the bins as objects.

One exception is \cite{CKKP} where each ball has its own maximal acceptable
congestion in each bin: this is exactly like our description of top-$n$-
guarantees (section 2.2), with the difference that the caps are exogenous in 
\cite{CKKP} so that it may not be feasible to assign all balls. The paper
shows the complexity of computing the maximal number of balls we can assign
and evaluates the price of anarchy of the corresponding free mobility
equilibrium.

\section{the ordinal model: weighted and unweighted}

We have $m$ posts denoted $a,b,\cdots $, and their set is $A$. Each agent $i$
in the finite set $N$ of cardinality $n$ must be assigned to some post $a$
in $A$.

An \textit{assignment} of agents to posts is a (quasi) partition $%
P=(S_{a})_{a\in A}$ of $N$ where $S_{a}$ is the set of agents assigned to
post $a$; the sets $S_{a},S_{b}$ are mutually disjoint and at most $m-1$ of
them can be empty. The set of all assignments is $\mathcal{P}(A,N)$, or
simply $\mathcal{P}$.

\paragraph{anonymous congestion: the unweighted model}

Each agent adds one unit of congestion at $a$. Given an assignment $P\in 
\mathcal{P}(A,N)$ the congestion at post $a$ is $s_{a}=|S_{a}|$ (the
cardinality of $S_{a}$); we call $s=(s_{a})_{a\in A}$ the \textit{congestion
profile} of $P$.

With the notation $[q]$ for the interval $\{1,\cdots ,q\}$ in $%
%TCIMACRO{\U{2115} }%
%BeginExpansion
\mathbb{N}
%EndExpansion
$, agent $i$'s (transitive and complete) preference relation $\succeq _{i}$
bears on the set $\mathcal{F}=A\times \lbrack n]$ of feasible allocations $%
(a,s_{a})$: $i$ is assigned to post $a$ where the congestion is $s_{a}$.
Preferences are strictly decreasing in $s_{a}$ and otherwise arbitrary. The
choice of $A,N$ and a preference profile $\succeq =(\succeq _{i})_{i\in N}$
defines a congestion problem $(A,N,\succeq )$.

\paragraph{agent-specific congestion: the weighted model}

Each agent $i$ brings the amount $w_{i}$ of congestion to her assigned post: 
$w_{i}$ is a \textit{strictly positive} real number and the total congestion
is $W=\sum_{i\in N}w_{i}$. For any subset $S$ of $N$ we use the notation $%
w_{S}=\sum_{j\in S}w_{j}$.

Given an assignment $P\in \mathcal{P}(A,N)$, the congestion profile is now $%
s=(w_{S_{a}})_{a\in A}$. The set of feasible allocations $(a,s_{a})$ becomes
agent-specific: $\mathcal{F}_{i}=A\times \lbrack w_{i},W]$; agent $i$ has
continuous preferences $\succeq _{i}$ over $\mathcal{F}_{i}$ strictly
decreasing in the congestion coordinate. A congestion problem is a 4-tuple $%
(A,N,w,\succeq )$.

The discussion of both models with ordinal preferences is the subject of
sections 3 to 5.

\section{the canonical guarantee: anonymous congestion}

Ex ante fairness provides to each agent $i$ a worst case welfare level that
only depends, besides their own preferences, upon the number of other
participants: our agent is essentially guaranteed one of their $n$ best out
of the $m\times n$ allocations in $\mathcal{F}$.

Fixing agent $i$ with preferences $\succeq _{i}$ on $\mathcal{F}$ a $n$-%
\textit{prefix }of $\succeq _{i}$ is a subset of $n$ allocations such that
each one is weakly preferred to all $(m-1)n$ other allocations. It is unique
if $\succeq _{i}$ is strict. Welfare decreases strictly in congestion
therefore if $z=(a,s)$ is in a certain $n$-prefix, so does $(a,s^{\prime })$
for all $s^{\prime }\leq s-1$. So we can represent a $n$-prefix of $\succeq
_{i}$ by a vector $\lambda _{i}=(\lambda _{ia})_{a\in A}$ such that%
\begin{equation}
\lambda _{ia}\in 
%TCIMACRO{\U{2115} }%
%BeginExpansion
\mathbb{N}
%EndExpansion
\cup \{0\}\text{ for all }a\text{, and}\sum_{a\in A}\lambda _{ia}=n
\label{1}
\end{equation}%
The $n$-prefix is the union over all $a$ of the sets $\{a\}\times \lbrack
\lambda _{ia}]$ in $\mathcal{F}$, with the convention $\{a\}\times \lbrack
0]=\varnothing $. We write $\Delta ^{%
%TCIMACRO{\U{2115} }%
%BeginExpansion
\mathbb{N}
%EndExpansion
}(A;n)$ for the set of $n$-prefixes defined by (\ref{1}).

For instance the $n$-prefix $\lambda _{ia}=n,\lambda _{ib}=0$ for $b\neq a$
means that $i$ (weakly) prefers post $a$ to any other post $x$, no matter
how congested $x$ and $a$ are. Another example has $\lambda _{ia}=\lfloor 
\frac{n}{m}\rfloor $ or $\lceil \frac{n}{m}\rceil $ for all $a$ (where $%
\lfloor x\rfloor $ and $\lceil x\rceil $ are the smallest and largest\
integer bounded above and below by $x$): this agent's priority is to
minimise the congestion level at whatever post they are assigned (up to the
necessary rounding up or down).

When indifferences in the preference $\succeq _{i}$ allow several $n$%
-prefixes $\lambda _{i}$, it is easy to check that any two of these differ
by at most $1$ in each coordinate $\lambda _{ia}$.

Under anonymous congestion we define Ex Ante fairness as the guarantee that
each agent's allocation is in one of their $n$-prefixes.\smallskip

\textbf{Definition 1} \textit{Given a problem }$(A,N,\succeq )$\textit{\ and
a profile }$\lambda =(\lambda _{i})_{i\in N}$ \textit{of corresponding }$n$%
\textit{-prefixes (one }$\lambda _{i}\in \Delta ^{%
%TCIMACRO{\U{2115} }%
%BeginExpansion
\mathbb{N}
%EndExpansion
}(A;n)$ for each preferences $\succeq _{i}$\textit{) the assignment }$P\in 
\mathcal{P}$\textit{\ is top-}$n$\textit{-fair iff}%
\begin{equation}
s_{a}\leq \lambda _{ia}\text{ for all }a\in A\text{ and }i\in S_{a}
\label{2}
\end{equation}

We write $\mathcal{P}(\lambda )$ for the set of top-$n$-fair assignments and 
$\mathcal{C(\lambda })$ for the set of congestion profiles $s$ when $P$
varies in $\mathcal{P}(\lambda )$.\smallskip

\textbf{Proposition 1 }\textit{In the problem }$(A,N,\succeq )$\textit{\
there exists at least one top-}$n$\textit{-fair assignment }$P$ \textit{for
any profile }$\lambda $ \textit{of corresponding }$n$\textit{%
-prefixes.\smallskip }

\textbf{Proof }We use a simple greedy algorithm and an induction argument on 
$n$.

If for some post $a$ we have $\lambda _{ia}=0$ for all $i$ we set $%
S_{a}=\varnothing $ and it remains to prove the claim on the residual
problem $(A\diagdown \{a\},N,\lambda )$. We clean up in this way all posts
that nobody accepts and we assume from now on that $\max_{i}\lambda
_{ia}\geq 1$ for all $a$.

Clearly if the result holds for a given $n$, the existence of a top-$n$-fair
assignment holds as well for all profiles $\widetilde{\lambda }$ weakly
larger than $\lambda \in \Delta ^{%
%TCIMACRO{\U{2115} }%
%BeginExpansion
\mathbb{N}
%EndExpansion
}(A;n)$ in all coordinates.

Pick any post\ $a$\ and order the caps $\lambda _{ia},i\in N$\ as $\lambda
^{\ast 1}\geq \lambda ^{\ast 2}\geq \cdots \geq \lambda ^{\ast n}$. Write $%
\widetilde{k}$ for the largest $k$ s.t. $\lambda ^{\ast k}\geq k$ (well
defined because $\lambda ^{\ast 1}\geq 1$) and pick a $\widetilde{k}$-prefix 
$S_{a}$ of the following ordering of $N$: $i\dashv ^{a}j\Leftrightarrow
\lambda _{ia}\geq $ $\lambda _{ja}$ of $N$. Then $\{\lambda _{ia}\}_{i\in
S_{a}}=\{\lambda ^{\ast k}\}_{1\leq k\leq \widetilde{k}}$ and $\lambda
_{ja}\leq \widetilde{k}$ for each $j\in N\diagdown S_{a}$ by definition of $%
\widetilde{k}$.

Assigning $S_{a}$ to $a$ meets inequalities (\ref{2}) for $a$ and in the
residual problem $(A\diagdown \{a\},N\diagdown S_{a},\lambda )$ we have $%
\sum_{b\in A\diagdown \{a\}}\lambda _{jb}\geq n-\widetilde{k}=|N\diagdown
S_{a}|$ for all $j$. The induction assumption on $n$ concludes the proof. $%
\blacksquare \smallskip $

Importantly the top-$n$-guarantee is \textit{maximal}: if all preferences $%
\succeq _{i}$ coincide then in any assignment $P$ and for any common $n$%
-prefix $\lambda _{0}$, at least one agent gets at $P$ their least preferred
allocation in $\lambda _{0}$: therefore the top-$n$-guarantee cannot be
improved.\smallskip

We illustrate the power of top-$n$-fairness (\ref{2}) in three examples. In
the first two $\mathcal{C(\lambda })$ is a singleton and we show the simple
way to recognise this fact.\smallskip

\textbf{Example 1: two posts. }\textit{We check that there is a single top-}$%
n$\textit{-fair congestion profile:} $|\mathcal{C}(\lambda )|=1$. Set $%
A=\{a,b\}$ and pick a profile $\lambda =(\lambda _{i})_{i\in N}$ of $n$%
-prefixes. Label the agents from $1$ to $n$ so that $\lambda _{ia}$
decreases (weakly) with $i$, while $\lambda _{ib}$ increases (weakly).
Keeping in mind $\lambda _{ia}+\lambda _{ib}=n$, the integer $\widetilde{k}$
in the proof of Lemma 1 is defined by the inequalities $\lambda _{\widetilde{%
k}a}\geq \widetilde{k}\geq \lambda _{(\widetilde{k}+1)a}\Longleftrightarrow
\lambda _{\widetilde{k}b}\leq n-\widetilde{k}\leq \lambda _{(\widetilde{k}%
+1)b}.$

We see that the maximal congestion at $a$ compatible with top-$n$-fairness
is $\widetilde{k}$ while at $b$ it is $n-\widetilde{k}$: therefore $(%
\widetilde{k},n-\widetilde{k})$ is the only top-$n$-fair congestion.
Moreover if $\lambda _{\widetilde{k}}\neq \lambda _{(\widetilde{k}+1)}$
there is also a single top-$n$-fair assignment.\smallskip

\textbf{Example 2: }\textit{Here }$m=4,n=18$\textit{, and\ there is a single
top-}$n$\textit{-fair assignment: }$|\mathcal{P}(\lambda )|=1$.

Agents are of five types labeled $\alpha $ to $\varepsilon $. Agents of a
given type have the same unique $18$-prefix but not necessarily identical
preferences, and the profile $\lambda $ is%
\begin{equation*}
\begin{array}{ccccc}
& a & b & c & d \\ 
\alpha \alpha \alpha & 3 & 2 & 10 & 3 \\ 
\beta \beta \beta \beta & 7 & 1 & 7 & 3 \\ 
\gamma \gamma & 3 & 8 & 3 & 4 \\ 
\delta \delta \delta \delta \delta & 4 & 2 & 9 & 3 \\ 
\varepsilon \varepsilon \varepsilon \varepsilon & 4 & 0 & 7 & 7%
\end{array}%
\end{equation*}

Check that $\mathcal{P}(\lambda )$ contains just one assignment $P$%
\begin{equation}
P:%
\begin{array}{cccc}
a & b & c & d \\ 
\beta \beta \beta \beta & \gamma \gamma & \alpha \alpha \alpha \delta \delta
\delta \delta \delta & \varepsilon \varepsilon \varepsilon \varepsilon%
\end{array}
\label{19}
\end{equation}%
We can fit top-$18$-fairly at most $4$ agents at post $a$; at most $2$ at
post $b$; at most $8$ at post $c$; at most $4$ at post $d$. As $4+2+8+4=18$
the only top-$18$-fair congestion profile is $(4,2,8,4)$; next we see that
the only way to fit $8$ agents at $c$ is with all $\alpha $-s and $\delta $%
-s; then we must assign the $\gamma $-s to $b$ and $\mathcal{P}(\lambda
)=\{P\}$ follows.

More generally write $c^{mx}(a;\lambda )$ for the maximal number of agents
we can fit $\lambda $-fairly at post $a$. Lemma 1 implies $\sum_{a\in
A}c^{mx}(a;\lambda )\geq n$ for any $\lambda $. If $\sum_{a\in
A}c^{mx}(a;\lambda )=n$ then all top-$n$-fair assignments have the same
congestion profile $s_{a}=c^{mx}(a;\lambda )$ for all $a$. The converse
property holds as well (we omit the easy proof):%
\begin{equation}
|\mathcal{C(\lambda })|=1\Longleftrightarrow \sum_{a\in A}c^{mx}(a;\lambda
)=n  \label{29}
\end{equation}

Examples where$|\mathcal{P}(\lambda )|>|\mathcal{C}(\lambda )|=1$ abound, e.
g. Example 1 if $\lambda _{\widetilde{k}}\neq \lambda _{(\widetilde{k}+1)}$,
or our next example.

\textbf{Example 3: }\textit{Here}\textbf{\ }$m=3,n=12$ and $|\mathcal{C}%
(\lambda )|=15,|\mathcal{P}(\lambda )|=953$.

Agents are of two types, six in each type with identical $12$-prefixes:%
\begin{equation*}
\begin{array}{cccc}
& a & b & c \\ 
\alpha \alpha \alpha \alpha \alpha \alpha & 6 & 4 & 2 \\ 
\beta \beta \beta \beta \beta \beta & 2 & 4 & 6%
\end{array}%
\end{equation*}

There are three top-$12$-fair congestion profiles respecting the symmetry
between the two types:%
\begin{equation}
\begin{array}{c}
\\ 
P_{1} \\ 
P_{2} \\ 
P_{3}%
\end{array}%
\begin{array}{ccc}
a & b & c \\ 
\alpha \alpha \alpha \alpha \alpha \alpha & 0 & \beta \beta \beta \beta
\beta \beta \\ 
\alpha \alpha \alpha \alpha \alpha & \alpha \beta & \beta \beta \beta \beta
\beta \\ 
\alpha \alpha \alpha \alpha & \alpha \alpha \beta \beta & \beta \beta \beta
\beta%
\end{array}
\label{43}
\end{equation}

In addition $\mathcal{C}(\lambda )$ contains six asymmetric congestion
profiles $(s_{a},s_{b},s_{c})=(6,1,5)$, $(6,2,4)$, $(6,3,3)$, $(6,4,2)$, $%
(5,4,3)$, $(5,3,4)$, and six more by exchanging the role of $\alpha $ and $%
\beta $.

Count next the top-$n$-fair assignments: in (\ref{43}) $P_{1}$ allows just
one assignment, but $P_{2}$ and $P_{3}$ allow respectively $36$ and $225$
assignments by permuting agents within their types.\footnote{%
For $P_{3}$ at post $b$ fifteen choices of the $\beta $-s and fifteen of the 
$\alpha $-s .} Taking all profiles of $\mathcal{C}(\lambda )$ into account
gives $|\mathcal{P}(\lambda )|=953$.

Depending on individual preferences, some of the top-$12$-fair assignments
(but not all\footnote{%
If $P\in \mathcal{P}(\lambda )$ is efficient within $\mathcal{P}(\lambda )$,
it is clearly efficient within the entire feasible set.}) may be inefficient
(Pareto dominated). For instance if in $P_{3}$ four $\alpha $-agents prefer $%
(a,4)$ to $(b,4)$ and the last two have the opposite preference then only
one of the $225$ corresponding assignments of the $\alpha $-s between posts $%
a$ and $b$ is efficient.

\paragraph{two strategic interpretations of top-$n$-fairness}

Consider the direct revelation mechanism where agents report independently
their preferences and the mechanism implements a reportedly top-$n$-fair
assignment. Then every Nash equilibrium of such mechanism must be top-$n$%
-fair because the truthful report ensures this. The same is true in any
mechanism in which the last stage offers to each agent the chance to claim a
top-$n$-fair allocation.

For the second interpretation we suppose that agent $i$, clueless about
other agents' preferences at the interim stage, wants to maximise their
worst case welfare: reporting a truthful $n$-prefix of $\succeq _{i}$\
ensures this and no other report does. Indeed in the adversarial case where
everyone else reports the same $\lambda _{i}$ an assignment rule treating
equals as equally as possible can give agent $i$\ anyone of the allocations $%
(a,\lambda _{ia})$, in particular a worst one in this set. Of course, any
information about other participants' preferences can be used to advantage
by our agent but at some risk.

\paragraph{free mobility equilibria and top-$n$-fairness}

In the FM normal form game each agent picks their post $a$ in $A$ and
consumes $(a,s_{a})$ where $s_{a}-1$ is the number of other agents who chose 
$a$. We check that every Nash equilibrium is a top-$n$-fair assignment. Fix
a profile of strategies $x=(x_{i})_{i\in N}\in A^{N}$ and write $s(a|x)$ for
the resulting congestion at post $a$. The equilibrium property for $i$ is $%
(x_{i},s(x_{i}|x))\succeq _{i}(a,s(a|x)+1)$ for all $a,a\neq x_{i}$: at such
post $a$ the only allocations $i$ may strictly prefer to $(x_{i},s(x_{i}|x))$
are in $\{a\}\times \lbrack s(a|x)]$ so their number is at most $s(a|x)$;
adding to those the allocations in $\{x_{i}\}\times \lbrack s(x_{i}|x)-1]$
that our agent definitely prefers to $(x_{i},s(x_{i}|x))$, we see that at
most $(n-1)$ allocations can beat $(x_{i},s(x_{i}|x))$ and conclude that the
latter belongs to at least one $n$-prefix of agent $i$.

Recall that the FM game has at least one Nash equilibrium assignment (\cite%
{Mil}). Clearly these assignments can be a strict subset of $\mathcal{P}%
(\lambda )$; e. g. in Example 3 $P_{1}$ is a FM eq. assignment only if $%
(a,6)\succeq _{\alpha }(b,1)$ and $(c,6)\succeq _{\beta }(b,1)$.

\section{the canonical guarantee: weighted congestion}

In the interim stage at which we define Ex Ante fairness, agent $i$ knows
their own weight $w_{i}$ and the total congestion $W$, but neither the
number of other agents filling the congestion $W-w_{i}$ or their individual
weights. The welfare level guaranteed to $i$ depends only upon these two
variables and $i$'s preferences. This definition does not directly
generalise that of guarantees under anonymous congestion because adversarial
situations allows for any finite split of the total weight $W-w_{i}$ of
other participants.

The (Lebesgue measure) size of the set of agent $i$'s allocations $\mathcal{F%
}_{i}=A\times \lbrack w_{i},W]$ is $m(W-w_{i})$: we show below that agent $i$
can guarantee one of their top $\frac{1}{m}$ allocations in $\mathcal{F}_{i}$%
, i. e., their best subset of size $(W-w_{i})$. Exactly like under anonymous
congestion where out of $m\times n$ allocations in $\mathcal{F}$, agent $i$%
's $n$-prefixes are the top $\frac{1}{m}$-th quantiles of $i$'s preferences.

The description of agent $i$'s $(W-w_{i})$-prefix is more subtle than for $n$%
-prefixes, but on the bright side this prefix is unique for each preference $%
\succeq _{i}$. With the notation $[[z]]$ for the number of strictly positive
coordinates of $z\in 
%TCIMACRO{\U{211d} }%
%BeginExpansion
\mathbb{R}
%EndExpansion
_{+}^{A}$, a $(W-w_{i})$-prefix is described by a vector $\lambda
_{i}=(\lambda _{ia})_{a\in A}$ such that%
\begin{equation}
\text{for each }a:\lambda _{ia}=0\text{ or }w_{i}\leq \lambda _{ia}\leq W%
\text{, and }\sum_{a\in A}\lambda _{ia}=W+([[\lambda _{i}]]-1)w_{i}
\label{6}
\end{equation}

The $(W-w_{i})$-prefix captured by the vector $\lambda _{i}$ above is as
follows. If $\lambda _{ia}\in \lbrack w_{i},W]$ it contains all allocations
in $\{a\}\times \lbrack w_{i},\lambda _{ia}]$. In particular $\lambda
_{ia}=w_{i}$ means that agent $i$ can be at post $a$ only if she is alone
there. If $\lambda _{ia}=0$ agent $i$ cannot be assigned to post $a$.

Notice that changing $\lambda _{ia}$ from $0$ to $w_{i}$ adds $w_{i}$ to
both sides of (\ref{6}), therefore ruling out post $a$ out or accepting it
only if $i$ is alone there has no impact on the equality constraint in (\ref%
{6}). For instance if $\lambda _{ia}=W$ our agent can still choose to accept
any subset of the other posts provided she is alone there.

Let $B$ be the (strict and possibly empty) subset $B$ of the posts such that 
$\lambda _{i}=0$; the total size of the allocations in $(A\diagdown B)\times
\lbrack w_{i},W]$ that $\lambda _{i}$ allows is%
\begin{equation*}
\sum_{a\in A\diagdown B}(\lambda _{ia}-w_{i})=\sum_{a\in A}\lambda
_{ia}-(m-|B|)w_{i}=W-w_{i}
\end{equation*}%
therefore $\lambda _{i}$ cuts a subset of size $(W-w_{i})$ in $\mathcal{F}%
_{i}$ as announced two paragraphs before (\ref{6}). Moreover for any
preference $\succeq _{i}$the top $\frac{1}{m}$ quantile in $\mathcal{F}%
_{i}=A\times \lbrack w_{i},W]$ is unique: this is clear by continuity of
preferences and strictness in each $\{a\}\times \lbrack w_{i},W]$. Then we
set $\lambda _{ia}=0$ if and only if this set does not contain any
allocation at post $a$ and $\lambda _{i}$ is uniquely defined as
well.\smallskip

\textbf{Definition 2} \textit{In the problem }$(A,N,\succeq ,w)$ \textit{%
where }$\lambda _{i}$\textit{\ meeting (\ref{6}) is the }$(W-w_{i})$\textit{%
-prefix of }$\succeq _{i}$\textit{, an assignment }$P$\textit{\ is top-}$%
\frac{1}{m}$\textit{-fair\ if and only if}%
\begin{equation*}
\sum_{j\in S_{a}}w_{j}\leq \lambda _{ia}\text{ for all }a\in A\text{ and all 
}i\in S_{a}
\end{equation*}

\textbf{Proposition 2 }\textit{In any problem }$(A,N,\succeq ,w)$\textit{\
there exists at least one top-}$\frac{1}{m}$\textit{-fair\ assignment }$P$%
\textit{.\smallskip }

Proof in the Appendix section 8.1. There we also state and prove that the
canonical guarantee is maximal, just like when congestion is
anonymous.\smallskip

Finding the top-$\frac{1}{m}$-fair assignments is more difficult with
weighted rather than unweighted congestion. This is already clear with two
posts.

\textbf{Example 4: }\textit{Two posts and three agents, }$m=2,n=3$\textit{,
and}\textbf{\ }$W=10$. The weights $w_{i}$ and $(W-w_{i})$-prefixes are%
\begin{equation*}
\begin{array}{cccc}
& a & b & w \\ 
\alpha & 10 & 6 & 6 \\ 
\beta & 6 & 6 & 2 \\ 
\gamma & 8 & 4 & 2%
\end{array}%
\end{equation*}%
Here $\alpha $'s feasible set is $\mathcal{F}_{\alpha }=\{a\}\times \lbrack
6,10]\cup \{b\}\times \lbrack 6,10]$ and its $4$-prefix $\{a\}\times \lbrack
6,10]\cup \{(b,6)\}$; similarly $\gamma $ accepts allocations in $%
\{a\}\times \lbrack 2,8]\cup \{b\}\times \lbrack 2,4]$.

Suppose every agent $i$'s preferences over allocations are represented by
the \textit{slack} \textit{utilities}:\footnote{%
This simple utility function derived from just its individual prefix is
frequently used in the rest of the paper, both with ordinal preferences and
cardinal \ utilities.}%
\begin{equation}
u_{i}(x,w_{S_{x}})=\lambda _{ix}-w_{S_{x}}\text{ for all }x  \label{16}
\end{equation}%
We find three top-$\frac{1}{2}$-fair assignments with very different
congestion and utilities%
\begin{equation*}
\text{assignments: }%
\begin{array}{ccc}
& a & b \\ 
P_{1} & \alpha & \beta \gamma \\ 
P_{2} & \alpha \gamma & \beta \\ 
P_{3} & \beta \gamma & \alpha%
\end{array}%
\text{ utilities: }%
\begin{array}{cccc}
& \alpha & \beta & \gamma \\ 
P_{1} & 4 & 2 & 0 \\ 
P_{2} & 2 & 4 & 0 \\ 
P_{3} & 0 & 2 & 4%
\end{array}%
\end{equation*}%
Recall that with anonymous congestion and strict preferences (Example 1
section 3) two-post problems have a unique top-$n$-assignment.

A Free Mobility equilibrium with weighted congestion may have no equilibrium
(see an example in section 8 of \cite{Mil}) but if it does, the
corresponding assignment is top-$\frac{1}{m}$-fair. Fix such an assignment $%
P $\ where $(a,w_{S_{a}})$\ is agent $i$'s equilibrium allocation. To $%
(a,w_{S_{a}})$\ our agent prefers the allocations $\{a\}\times \lbrack
w_{i},w_{S_{a}}]$\ and possibly those in $\{x\}\times \lbrack
w_{i},w_{S_{x}}+w_{i}]$: the total length of those sets is exactly $W-w_{i}$%
, which proves the claim.

\section{deterministic competitive assignments}

\subsection{under anonymous congestion}

We use the notation $z\vee y=\max \{z,y\}$.\smallskip

\textbf{Definition 3} \textit{In problem }$(A,N,\succeq )$ \textit{the
assignment }$P$\textit{\ is competitive (Comp) iff}%
\begin{equation}
\text{for all }a\in A\text{ and }i\in S_{a}:(a,s_{a})\succeq
_{i}(x,s_{x}\vee 1)\text{ for all }x\in A  \label{18}
\end{equation}

If all posts are occupied ($s_{x}\geq 1$ for all $x$) property (\ref{18}) is
just envy-freeness: switching my allocation to yours is a virtual move that
does not affect the congestion at your post (unlike an actual move in the FM
game). Interpreting the congestion $s_{x}$ at post $x$ as the
\textquotedblleft price\textquotedblright\ of consuming $x$, my competitive
demand is the allocation $(x,s_{x})$ when I assume that the price vector is
fixed.

Property (\ref{18}) assigns a price of $1$ to an empty post $x$ and insists
that our agent does not want to move there either: this is important because
an assignment where all agents share a single post $a$ is automatically
envy-free, but can be the absolute worst assignment for everyone.\smallskip

\textbf{Proposition 3 }\textit{Fix a} \textit{problem }$(A,N,\succeq )$.

\noindent $i)$ \textit{All competitive assignments have the same congestion
profile (except possibly at some posts occupied by at most one agent), and
the same welfare profile.}

\noindent $ii)$ \textit{A competitive assignment is weakly efficient, and
efficient if preferences are strict and/or if all posts are occupied.}

\noindent $iii)$ \textit{A competitive assignment is top-}$n$\textit{-fair}%
.\smallskip

\textbf{Proof }At the congestion profile $s\in \Delta ^{%
%TCIMACRO{\U{2115} }%
%BeginExpansion
\mathbb{N}
%EndExpansion
}(A;n)$ ((\ref{1})) we define agent $i$'s \textit{competitive} \textit{demand%
} as%
\begin{equation}
D(i;s)=\{a|(a,s_{a}\vee 1)\succeq _{i}(x,s_{x}\vee 1)\text{ for all }x\in A\}
\label{22}
\end{equation}%
The assignment $P$ is competitive if and only if $a\in D(i;s)$ whenever $%
i\in S_{a}$.\smallskip

\noindent \textit{Statement }$i)$ \textit{Unique congestion. }We fix two
different congestion profiles $s,s^{\ast }$ coming from the competitive
assignments $P=(S_{x})_{x\in A}$ and $P^{\ast }=(S_{x}^{\ast })_{x\in A}$.

Define the set $A^{\ast }=\{a\in A:s_{a}^{\ast }\vee 1>s_{a}\vee 1\}$ and
assume that $A^{\ast }$ is non empty, which will lead to a contradiction.
Note that in $A^{\ast }$ we have $s_{a}^{\ast }>s_{a},1$.

Fixing an agent $i$ we claim that if $D(i;s^{\ast })$ intersects $A^{\ast }$
then $D(i;s)$ must be a subset of $A^{\ast }$. If the claim fails there is
some $b\in D(i;s)$ outside $A^{\ast }$ such that for all $a$ in $A^{\ast }$: 
$(b,s_{b}\vee 1)\succeq _{i}(a,s_{a}\vee 1)\succ _{i}(a,s_{a}^{\ast })$
(strict preference because $s_{a}^{\ast }>s_{a},1$). By the choice of $b$ we
also have $s_{b}\vee 1\geq s_{b}^{\ast }\vee 1$; therefore $(b,s_{b}^{\ast
}\vee 1)\succeq _{i}(b,s_{b}\vee 1)\succ _{i}(a,s_{a}^{\ast }),$ and as $a$
was arbitrary\ in $A^{\ast }$ it follows that $D(i;s^{\ast })$ cannot
intersect $A^{\ast }$, contradiction.

Now for each $i\in {\Large \cup }_{a\in A^{\ast }}S_{a}^{\ast }$ the claim
says $D(i;s)\subseteq A^{\ast }$ therefore $i$ is assigned to $A^{\ast }$ by 
$P$. This implies $\sum_{a\in A^{\ast }}s_{a}\geq \sum_{a\in A^{\ast
}}s_{a}^{\ast }$, contradicting $s_{a}^{\ast }>s_{a}$ in $A^{\ast }$. We
conclude that $A^{\ast }$ is empty.

So $P,P^{\ast }$ must be such that $s_{a}^{\ast }\vee 1=s_{a}\vee 1$ for all 
$a$: SO $s_{a}^{\ast }\neq s_{a}$ can only happen when $\{s_{a}^{\ast
},s_{a}\}=\{0,1\}$, as claimed in statement $i)$. Moreover the competitive
demands at $s$ and $s^{\ast }$ coincide.

A simple example with multiple competitive congestion profiles has all
agents except $1$ and $2$ refusing the three posts $a,b,c$, while $1$ and $2$
refuse all but $a,b,c$ and they are indifferent between $(a,1),(b,1)$ and $%
(c,1)$: combining a competitive sub-assignment of $N\diagdown \{1,2\}$ to $%
A\diagdown \{a,b,c\}$ with any assignment where agents $1$,$2$ occupy two of 
$a,b,c$ is competitive in the full problem.

\noindent\ \textit{Unique welfare. }We fix an agent $i$ in $S_{a}$ and $%
S_{b}^{\ast }$ and show that $i$ is indifferent between the two assignments.
From $s_{a},s_{b}^{\ast }\geq 1$ and Definition 3 we have%
\begin{equation}
(a,s_{a})\succeq _{i}(b,s_{b}\vee 1)\text{ and }(b,s_{b}^{\ast })\succeq
_{i}(a,s_{a}^{\ast }\vee 1)  \label{10}
\end{equation}%
If $s_{a}=s_{a}^{\ast }$ and $s_{b}=s_{b}^{\ast }$ we are done. If $%
s_{a}\neq s_{a}^{\ast }$ and $s_{b}=s_{b}^{\ast }$ the equality $s_{a}^{\ast
}\vee 1=s_{a}\vee 1$ implies $s_{a}=1>0=s_{a}^{\ast }$; then $s_{b}^{\ast
}\geq 1$ and (\ref{10}) give $(a,1)\succeq _{i}(b,s_{b})=(b,s_{b}^{\ast
})\succeq _{i}(a,1)$ as desired. The last subcase $s_{a}\neq s_{a}^{\ast }$
and $s_{b}\neq s_{b}^{\ast }$ is just as easy.\smallskip

\noindent \textit{Statement }$ii)$ \textit{Efficiency. }Assume, to the
contrary that $P=(S_{x})$ is competitive and Pareto inferior to $Q=(T_{x})$.
Say agent $i$,$\ $assigned to $a$ at $P$, is assigned to $b$ at $Q$ ($a,b$
are not necessarily distinct) and suppose that post $b$ is occupied at $P$: $%
s_{b}\geq 1$. Then by Comp and the weak Pareto improvement we have%
\begin{equation}
(b,s_{b})\preceq _{i}(a,s_{a})\preceq _{i}(b,t_{b})\Longrightarrow s_{b}\geq
t_{b}  \label{12}
\end{equation}%
and $s_{b}>t_{b}$ if agent $i$ improves strictly at $Q$.

If all posts are occupied at $P$ (\ref{12}) implies $s=t$ and we have a
contradiction. If instead some agent $i$ goes from $a$ at $P$ to $c$ at $Q$
and $c$ is empty at $P$, $s_{c}=0$, we have%
\begin{equation}
(c,1)\preceq _{i}(a,s_{a})\preceq _{i}(c,t_{c})\Longrightarrow t_{c}=1\text{
and }(a,s_{a})\simeq _{i}(c,1)  \label{8}
\end{equation}

This is a contradiction if preferences are strict, and we conclude again
that $P$ is Pareto optimal. If indifferences are possible we see that $i$ is
a weak Pareto optimum (not all agents benefit strictly).

The argument above explains how a competitive assignment can be Pareto
inferior: in the situation (\ref{8}) moving $i$ from $a$ to $c$ \textit{and
changing nothing else} is a Pareto improvement to a new assignment $%
\widetilde{P}$ where agents in $S_{a}\diagdown \{a\}$ benefit strictly while
all others are unaffected.

Note that $\widetilde{P}$ is not competitive if $s_{a}\geq 2$ because $i$
prefers strictly $(a,s_{a}-1)$ to $(a,s_{a})$ and therefore to $(c,1)$; but
if $s_{a}=1$ then $\widetilde{P}$ is competitive and welfare-wise
indifferent to $P$.

We have shown that Pareto improving a competitive assignment is only
possible if $\succeq _{i}$ is not strict and some post is
unoccupied.\smallskip

\noindent \textit{Statement }$iii)$ \textit{top}-$n$\textit{-fairness.} A FM
equilibrium is top-$n$-fair (section \textbf{3}). Now a competitive
assignment is clearly a FM equilibrium. $\blacksquare \smallskip $

Proposition 3 vindicates a competitive assignment as the essentially
single-valued, fair and efficient solution to the congested assignment
problem, precisely what we set out to discover. But it is as easy to find
problems where such assignment exists as where they don't. Recall that
deciding if one exists in a given problem is computable in polynomial time (%
\cite{CGW}). Deciding when existence is \textquotedblleft
likely\textquotedblright\ in some plausible domain of preferences is
therefore a feasible numerical and empirical project.

Start with an extreme example where $n=m$\ and every agent's $n$-prefix is $%
\lambda _{ia}=1$\ for all $a$: they insist on being alone at their assigned
post. An assignment is top-$n$-fair if and only if it is a one-to-one (non
congested) assignment. Therefore a competitive one exists only if we can
match each agent with one of their best posts. By contrast any assignment is
a FM equilibrium outcome.

In Example 2 (section 3) $P$ ((\ref{19})) is the the unique top-$18$-fair
assignment. Assume that the preferences of each agent $i$ with $n$-prefix $%
\lambda _{i}$ are described by the \textit{slack }utilities $u_{i}=\lambda
_{ia}-s_{a}$ as in (\ref{16}). Then it is easy to check that $P$ is
competitive.\footnote{%
At $P$ the congestion profile is $(a,4),(b,2),(c,8),(d,4)$ so the $\alpha $%
-s' utility at $c$ is $2$ and at most $0$ at any other post; the $\beta $-s'
utility at $a$ is $3$ and at most $1$ elsewhere; etc...} However, if the
preferences of some $\alpha $,say, change to $(b,2)\succ _{\alpha }(c,8)$
(compatible with the given $n$-prefixes) then $P$ is no longer competitive.

Turning to Example 3 (section 3) and maintaining the slack utilities
assumption, we find that \textit{none} of the hundreds of top-$12$%
-assignments is competitive: at $P_{1}$ in (\ref{43}) all agents want to
move to $b$; at $P_{2}$ all agents at $a$ or $c$ envy those at $b$, and the
reverse statement holds at $P_{3}$. Checking all asymmetric congestion
profiles is just as easy.

Our last two examples have three posts and six agents, $m=3,n=6$, and three
pairs of identical agents with a cyclical pattern in their preferences. We
use them to illustrate the welfare performance of competitive assignments
within the set of FM eq. assignments (to which they belong by Definition 3).

\textbf{Example 5.1} \textit{Where the competitive assignment Pareto
dominates two more FM equ. outcomes. }Each agent has a single $6$-prefix, a
single top-$n$-fair congestion profile $\mathcal{C}(\lambda )=\{(2,2,2)\}$
(by (\ref{29})) and ten top-$6$-fair assignments%
\begin{equation}
\begin{array}{cccc}
& a & b & c \\ 
\alpha \alpha & 3 & 2 & 1 \\ 
\beta \beta & 1 & 3 & 2 \\ 
\gamma \gamma & 2 & 1 & 3%
\end{array}%
\text{ \ \ \ \ }%
\begin{array}{c}
\\ 
P_{1} \\ 
P_{2} \\ 
P_{3}%
\end{array}%
\begin{array}{ccc}
a & b & c \\ 
\alpha \alpha & \beta \beta & \gamma \gamma \\ 
\gamma \gamma & \alpha \alpha & \beta \beta \\ 
\alpha \gamma & \alpha \beta & \beta \gamma%
\end{array}
\label{31}
\end{equation}
because $P_{3}$ defines eight assignments by permuting agents within types.

If preferences are described by the slack utilities (\ref{16}), only $P_{1}$
is competitive with utility $1$ for everyone while $P_{2},P_{3}$ are FM equ.
outcomes Pareto inferior to $P_{1}$.\footnote{%
Respectively with utility $0$ for everyone at $P_{2}$, or $0$ and $1$ in
each type at $P_{3}$.}\smallskip

\textbf{Example 5.2 }\textit{With a competitive assignment and other non
Pareto comparable FM eq. assignments. }The unique prefixes and preferences
are as follows:%
\begin{equation*}
\begin{array}{cccc}
& a & b & c \\ 
\alpha \alpha & 3 & 3 & 0 \\ 
\beta \beta & 3 & 0 & 3 \\ 
\gamma \gamma & 0 & 3 & 3%
\end{array}%
\text{ \ \ \ }%
\begin{array}{c}
\alpha :(a,2)\succ _{\alpha }(b,2)\succ _{\alpha }(b,3)\succ _{\alpha }(a,3)
\\ 
\beta :(c,2)\simeq _{\beta }(a,2) \\ 
\gamma :(c,1)\succ _{\gamma }(b,2)\succ _{\gamma }(b,3)\succ _{\gamma }(c,2)%
\end{array}%
\end{equation*}

The following competitive assignment $P_{1}$ is not welfare comparable to
the non-competitive FM equ. assignments $P_{2}$and $P_{3}$.\footnote{%
The $\alpha $-s prefer $P_{1}$ to $P_{2}$ and $P_{3}$ (at least weakly); the 
$\beta $-s prefer $P_{3}$ to $P_{1}$ (at least weakly) and are indifferent
between $P_{1}$ and $P_{2}$; the $\gamma $-s strictly prefer $P_{1}$ to $%
P_{3}$ and disagree when comparing $P_{1}$ to $P_{2}$.}%
\begin{equation*}
\begin{array}{c}
\\ 
P_{1} \\ 
P_{2} \\ 
P_{3}%
\end{array}%
\begin{array}{ccc}
a & b & c \\ 
\alpha \alpha & \gamma \gamma & \beta \beta \\ 
\beta \beta & \alpha \alpha \gamma & \gamma \\ 
\alpha \beta & \alpha \gamma \gamma & \beta%
\end{array}%
\end{equation*}

\subsection{under weighted congestion}

\textbf{Definition 4} \textit{In problem }$(A,N,\succeq ,w)$, t\textit{he
assignment }$P$\textit{\ is competitive (Comp) iff}%
\begin{equation*}
\text{for all}\mathit{\ }a\in A\mathit{\ }\text{and}\mathit{\ }i\in
S_{a}:(a,w_{S_{a}})\succeq _{i}(x,w_{S_{x}}\vee w_{i})\text{ for all}\mathit{%
\ }x\in A
\end{equation*}%
\textit{\ }

The interpretation is the same as before, except that the \textquotedblleft
congestion price\textquotedblright\ of a post is now agent-specific. As
before competitiveness implies top-fairness, but the critical uniqueness and
efficiency properties require some qualification.\smallskip

\textbf{Definition 5 }\textit{The assignment }$P$\textit{\ is crowded iff }$%
|S_{a}|\geq 2$\textit{\ or }$S_{a}=\varnothing $\textit{\ for all }$a$%
\textit{.}

\noindent \textit{The ordinal preferences }$\succeq _{i}$\textit{\ are
semi-strict iff for any distinct posts }$a,b$\textit{\ and any agent }$i$%
\textit{\ and }$S$\textit{\ s. t. }$i\in S\subseteq N$,\textit{\ agent }$i$ 
\textit{is not indifferent between }$(a,w_{i})$\textit{\ and }$(b,w_{S})$%
\textit{.}\smallskip

\textbf{Proposition 4: }\textit{Fix a} \textit{problem }$(A,N,\succeq ,w)$.

\noindent $i)$ \textit{If all preferences }$\succeq _{i}$\textit{\ are
semi-strict then all competitive assignments have the same congestion and
welfare profiles.} \textit{The same is true if at least one competitive
assignment is crowded.}

\noindent $ii)$ \textit{If all preferences are semi-strict a competitive
assignment is efficient.}

\noindent $iii)$ \textit{A competitive assignment is top-}$\frac{1}{m}$%
\textit{-fair\smallskip }

Proof in the Appendix section 8.2.

To ilIustrate the role of the semi-strictness assumption in statement $i)$
it is enough to check that in the two-post Example 4 section 4, $P_{1}$ and $%
P_{2}$ are both competitive assignments with different welfare and even
congestion profiles.

\section{competitive fractional assignments under anonymous congestion}

In the randomised version of the unweighted model we endow each agent $i$
with a cardinal vNM utility $u_{i}(a,s)$ over $A\times \lbrack n]$. Fixing
the post $a$ and some expected congestion $\sigma $ at $a$ we write $%
u_{i}(a,\sigma )$ for the linear interpolation of $u_{i}(a,s)$ between $%
\lfloor \sigma \rfloor $ and $\lceil \sigma \rceil $ (the two rounded up and
down values of $\sigma $):%
\begin{equation}
u_{i}(a,\sigma )=\frac{\lceil \sigma \rceil -\sigma }{\lceil \sigma \rceil
-\lfloor \sigma \rfloor }u_{i}(a,\lfloor \sigma \rfloor )+\frac{\sigma
-\lfloor \sigma \rfloor }{\lceil \sigma \rceil -\lfloor \sigma \rfloor }%
u_{i}(a,\lceil \sigma \rceil )\text{ for all non-integer }\sigma  \label{24}
\end{equation}%
So $u_{i}(a,\sigma )$ is continuous and strictly decreasing in $\sigma \in
\lbrack 1,n]$.

The competitive approach becomes operational because it selects in every
problem a unique competitive congestion profile and implements it by a
lottery over approximately competitive deterministic assignments. The
similar randomisation in the weighted congestion model still identifies an
essentially unique congestion profile, but its implementation is less
satisfactory. See subection 8.3 in the Appendix for details.\smallskip

For a finite set $Z$ the notation $\Delta ^{%
%TCIMACRO{\U{211d} }%
%BeginExpansion
\mathbb{R}
%EndExpansion
}(Z;y)$ is the simplex with non negative real coordinates in $Z$ adding up
to the real number $y$. The key variable of a random assignment is the
expected congestion $\sigma _{a}$ at each post $a$: the congestion profile $%
\sigma =(\sigma _{a})_{a\in A}$ varies in $\Delta ^{%
%TCIMACRO{\U{211d} }%
%BeginExpansion
\mathbb{R}
%EndExpansion
}(A;n)$. Note that $\sigma _{a}<1$ is possible: in this case the definitions
below make sure that each agent perceives a congestion of $1$ at $a$, so $%
u_{i}(a,\sigma )$ is \textit{not} defined if $\sigma \in \lbrack 0,1[$.
Moreover in Definition 7 describing the implementation of a random
congestion $\sigma $ by a lottery over deterministic assignments, post $a$'s
congestion at the latter assignments takes only the values $\lfloor \sigma
_{a}\rfloor $ and $\lceil \sigma _{a}\rceil $, so that equation (\ref{24})
is indeed the relevant vNM expected utility at $(a,\sigma )$.

As in the proof of Proposition 3 we define agent $i$'s \textit{competitive
demand} at the expected congestion profile $\sigma \in \Delta ^{%
%TCIMACRO{\U{211d} }%
%BeginExpansion
\mathbb{R}
%EndExpansion
}(A;n)$:%
\begin{equation}
D(u_{i},\sigma )=\arg \max_{x\in A}u_{i}(x,\sigma _{x}\vee 1)  \label{17}
\end{equation}%
This demand ignores the effect of agent $i$'s own presence at post $x$ but
correctly anticipates that the ex post congestion will round $\sigma _{x}$
up or down. Our key definition comes next.\smallskip

\textbf{Definition 6}: \textit{In the problem }$(A,N,u)$\textit{\ the
fractional congestion profile }$\sigma \in \Delta ^{%
%TCIMACRO{\U{211d} }%
%BeginExpansion
\mathbb{R}
%EndExpansion
}(A;n)$\textit{\ is competitive (Comp) iff}%
\begin{equation}
\sigma \in \sum_{i\in N}\Delta ^{%
%TCIMACRO{\U{211d} }%
%BeginExpansion
\mathbb{R}
%EndExpansion
}[D(u_{i},\sigma );1]  \label{13}
\end{equation}

Property (\ref{13}) means that we can achieve the congestion profile $\sigma
^{\ast }$ by assigning randomly each agent, with a well chosen probability
distribution, over the posts in her competitive demand. Writing this
probability distribution as $(\pi _{ia})$ we obtain a semi-stochastic matrix 
$\Pi =[\pi _{ia}]\in \lbrack 0,1]^{N\times A}$ (each row sums to $1$)
realising $\sigma $ and with competitive support:%
\begin{equation}
\sigma _{a}=\sum_{i\in N}\pi _{ia}\text{ for all }a\in A  \label{23}
\end{equation}%
\begin{equation}
\{\pi _{ia}>0\Longrightarrow a\in D(u_{i},\sigma )\}\text{ for all }i\in
N,a\in A  \label{25}
\end{equation}%
There may be more than one matrix $\Pi $ realising a given congestion
profile $\sigma $.\smallskip

\textbf{Lemma 1: }\textit{In every problem }$(A,N,u)$ \textit{there is a
unique competitive congestion profile }$\sigma ^{c}$\textit{, except
possibly at some posts where }$\sigma _{a}^{c}\leq 1$\textit{,\footnote{%
If $\sigma ,\sigma ^{\ast }$ are both competitive and $\sigma _{a}^{\ast
}\neq \sigma _{a}$ at some post $a$, then $\sigma _{a}^{\ast },\sigma _{a}$ $%
\leq 1$.} and a unique competitive demand}.\smallskip

\textbf{Proof }\textit{Existence}. It follows from the Kakutani fixed point
theorem applied to the convex compact valued correspondence $\Gamma (\sigma
)=\sum_{i\in N}\Delta ^{%
%TCIMACRO{\U{211d} }%
%BeginExpansion
\mathbb{R}
%EndExpansion
}(D(u_{i},\sigma );1)$ mapping $\Delta ^{%
%TCIMACRO{\U{211d} }%
%BeginExpansion
\mathbb{R}
%EndExpansion
}(A;n)$ into itself. To check that the graph of $\Gamma $ is closed
(implying that $\Gamma $ is upper-semi-continuous) we take a sequence $%
(\sigma ^{t},\tau ^{t})_{t=1}^{\infty }$ in $[\Delta ^{%
%TCIMACRO{\U{211d} }%
%BeginExpansion
\mathbb{R}
%EndExpansion
}(A;n)]^{2}$ converging to $(\sigma ,\tau )$ and s. t. $\tau ^{t}=\sum_{i\in
N}\tau _{i}^{t}$ and $\tau _{i}^{t}\in \Delta ^{%
%TCIMACRO{\U{211d} }%
%BeginExpansion
\mathbb{R}
%EndExpansion
}[D(u_{i},\sigma ^{t});1]$. We can find a subsequence such that all
sequences $\{\tau _{i}^{t}\}$ converge, and all sets $D(u_{i},\sigma ^{t})$
are constant in $t$, so that $\tau \in \Gamma (\sigma )$ as
desired.\smallskip

\noindent \textit{Uniqueness. }We adapt the proof of statement $i)$ in
Proposition 3. Assume $\sigma ,\sigma ^{\ast }$ are two different
competitive profiles with corresponding matrices $\Pi $ and $\Pi ^{\ast }$
in (\ref{23}), (\ref{25}). We set $A^{\ast }=\{a\in A|\sigma _{a}^{\ast
}\vee 1>\sigma _{a}\vee 1\}$ and check, exactly like in the deterministic
proof, that if an agent $i$ is s. t. $D(u_{i};\sigma ^{\ast })$ intersects $%
A^{\ast }$ at some $a$, then $D(u_{i};\sigma )$ must be a subset of $A^{\ast
}$.

Therefore for any $i$ s. t. $\sum_{a\in A^{\ast }}\pi _{ia}^{\ast }>0$ we
have $\sum_{a\in A^{\ast }}\pi _{ia}^{\ast }\leq 1=\sum_{a\in A^{\ast }}\pi
_{ia}$. Summing up over all agents this gives $\sum_{a\in A^{\ast }}\sigma
_{a}^{\ast }=\sum_{i\in N,a\in A^{\ast }}\pi _{ia}^{\ast }\leq \sum_{i\in
N,a\in A^{\ast }}\pi _{ia}=\sum_{a\in A^{\ast }}\sigma _{a}$, a
contradiction of the assumption $\sigma _{a}^{\ast }>\sigma _{a}$ in $%
A^{\ast }$. We conclude that $A^{\ast }$ is empty: $\sigma _{a}^{\ast }\vee
1\leq \sigma _{a}\vee 1$ for all $a$. The opposite inequality holds by
exchanging the roles of $\sigma $ and $\sigma ^{\ast }$, implying $\sigma
_{a}^{\ast }\vee 1=\sigma _{a}\vee 1$ and the desired conclusion that $%
\sigma _{a}^{\ast }\neq \sigma _{a}$ can only happen when both are at most $%
1 $.

Because an agent evaluates being assigned to a post with expected congestion
at most $1$ as being alone there (definition (\ref{17})) this implies $%
D(u_{i},\sigma )=D(u_{i},\sigma ^{\ast })$, therefore $\sigma $ and $\sigma
^{\ast }$ generate the same competitive demand. $\blacksquare \smallskip $

\textbf{Definition 7}: \textit{In the problem }$(A,N,u)$ \textit{the list }$%
(\{P^{k}\}_{k=1}^{K},\mathcal{L)}$\textit{\ of }$K$\textit{\ deterministic
assignments }$P^{k}$\textit{\ together with a lottery }$\mathcal{L}$ \textit{%
over} $[K]$\textit{, implements the competitive congestion profile }$\sigma
^{c}$ \textit{if, first, the expected congestion over these }$K$\textit{\
assignments is} $\sigma ^{c}$: $\mathbb{E}_{\mathcal{L}}(s_{a}^{k})=\sigma
_{a}^{c}$ \textit{for all }$a$\textit{; and second for all }$k\in \lbrack K]$%
, \textit{all }$i$\textit{,all }$a$\textit{\ and all }$k$ \textit{we have:}%
\begin{equation}
i\in S_{a}^{k}\Longrightarrow a\in D(u_{i},\sigma ^{c})\text{ and }%
s_{a}^{k}=\lfloor \sigma _{a}^{c}\rfloor \text{ \textit{or} }\lceil \sigma
_{a}^{c}\rceil  \label{14}
\end{equation}

Property (\ref{14}) says that anyone of the assignments $P^{k}$ is ex post
fair in the sense that each agent is at a post in his competitive demand
(based on the expected congestion $\sigma ^{c}$) and the actual congestion
at $P^{k}$ is an integer rounding of the latter.\smallskip

\textbf{Lemma 2: }\textit{In every problem }$(A,N,u)$\textit{\ we can
implement the competitive congestion profile }$\sigma ^{c}$ \textit{by
(typically several) lists }$(\{P^{k}\}_{k=1}^{K},\mathcal{L})$\textit{\ as
in Definition 4.}

\textbf{Corollary:}\textit{\ If }$\sigma ^{c}$\textit{\ is integer-valued
each assignment }$P^{k}$\textit{\ is competitive and implements }$\sigma
^{c} $\textit{.\smallskip }

\textbf{Proof }Define the set $\mathcal{S}$ of semi-stochastic matrices $\Pi 
$ s.t. for all $a$ and $i$: $\{\pi _{ia}>0\Longrightarrow a\in
D(u_{i},\sigma ^{c})\}$ and $\lfloor \sigma _{a}^{c}\rfloor \leq \sum_{i\in
N}\pi _{ia}\leq \lceil \sigma _{a}^{c}\rceil $. This set is a convex compact
polytope, non empty as it contains any matrix $\Pi ^{c}$ realising $\sigma
^{c}$. We claim that each extreme point $\Pi ^{k}$ of $\mathcal{S}$ is
deterministic, i. e., its entries are all integers or zero; so $\Pi ^{k}$ is
a deterministic assignment $P^{k}$ meeting (\ref{14}) and $\Pi ^{c}$ is a
convex combination of such extreme points: this gives us the desired
collection $P^{k}$ and lottery $\mathcal{L}$.

We prove the claim by contradiction. Pick an extreme point $\Pi $ of $%
\mathcal{S}$ and associate to $\Pi $ the bipartite graph $G$ on $N\times A$
containing the edge $ia$ iff $\pi _{ia}>0$. Extract from $G$ the subgraph $%
G_{0}$ of its fractional entries $ia$, i. e., $0<\pi _{ia}<1$, and let $F$
be the set of posts $a$ s.t. $\sum_{i\in N}\pi _{ia}$ is fractional (not an
integer or zero). If $F$ is non empty it contains some post $a$: at least
one edge $ia$ is in $G_{0}$; then at least one other edge $ib$ is in $G_{0}$
($\Pi $ is semi-stochastic at $i$); if $b\in F$ then we can add a small $%
\varepsilon $ to $\pi _{ia}$ and take away $\varepsilon $ from $\pi _{ib}$
without leaving $\mathcal{S}$, or vice versa: this contradicts the
extremality of $\Pi $, so $b$ is not in $F$ after all. But then there is
another edge $jb$ in $G_{0}$ because $\sum_{i\in N}\pi _{ib}$ is an integer,
and again there is some new edge $jc$ in $G_{0}$; if $c=a$ or $c\in F$ we
can as above perturb a little the entries of $\Pi $ in two opposite ways and
reach a contradiction. This construction must stop at a new post in $F$ or
cycle back to $a$: in both cases we can perturb the path or cycle in
opposite directions. The claim is proved and the proof is complete. $%
\blacksquare \smallskip $

Our main result refines the properties of the deterministic assignments $%
P^{k}$ selected ex post to implement the competitive congestion: any two
such assignments yield approximately identical utilities; and each $P^{k}$
shares (approximately) the properties of deterministic competitive
assignments in Proposition 1. We use for agent $i$'s approximation parameter
her worst utility loss from one additional unit of congestion $\delta
_{i}=\max_{(a,s)\in A\times \lbrack n]}\{u_{i}(a,s)-u_{i}(a,s+1)\}$. For
instance $\delta _{i}=1$ for a slack utility $u_{i}=\lambda _{ia}-s_{a}$%
.\smallskip

\textbf{Theorem 1 }\textit{Fix a problem }$(A,N,u)$\textit{, a list }$%
(\{P^{k}\}_{k=1}^{K},\mathcal{L})$ \textit{implementing the competitive
congestion }$\sigma ^{c}$ (Definition 6)\textit{\ and write }$%
U^{k}=(U_{i}^{k})_{i\in N}$\textit{\ the utility profile at assignment }$%
P^{k}$.

\noindent $i)$ \textit{Each }$P^{k}$\textit{\ is top-}$n$\textit{-fair\ up
to at most one unit of congestion}:\textit{\ for all }$i\in N$\textit{\
there is a }$n$\textit{-prefix }$\lambda _{i}$\textit{\ of }$u_{i}$\textit{\
such that }$i\in S_{a}^{k}\Longrightarrow s_{a}^{k}\leq \lambda _{ia}+1$.

\noindent $ii)$ \textit{Utilities at two assignments\ }$P^{k},P^{\ell }$ 
\textit{differ by at most }$2\delta _{i}$\textit{:} $|U_{i}^{k}-U_{i}^{\ell
}|\leq 2\delta _{i}$\textit{\ for all }$i$\textit{\ and all }$k,\ell \in
\lbrack K]$.

\noindent $iii)$\textit{\ Each }$P^{k}$ \textit{is }$2\delta _{i}$-\textit{%
competitive: }$U_{i}^{k}\geq u_{i}(x,s_{x}^{k}\vee 1)-2\delta _{i}$ \textit{%
for all }$i$\textit{\ and all }$x$\textit{.}

\noindent $iv)$\textit{\ Each assignment }$P^{k}$ \textit{is }$(2\delta
_{i}+\varepsilon )$-\textit{efficient for any} $\varepsilon >0$\textit{: no
assignment }$Q\in \mathcal{P}(A,N)$\textit{\ Pareto dominates the profile }$%
(U_{i}^{k}+2\delta _{i}+\varepsilon )_{i\in N}$.

\textit{If in addition\ }$\sigma _{a}^{c}\geq 1$\textit{\ for all }$a$%
\textit{, we can say more. Let }$U_{i}^{c}$\textit{\ be the value }$%
u_{i}(a,\sigma _{a}^{c})$\textit{\ common to all }$a\in D(u_{i},\sigma ^{c})$%
\textit{, then the profile }$(U_{i}^{c})_{i\in N}$\textit{\ is efficient%
\footnote{%
Not dominated by the utility profile of any \textit{assignment }$Q\in 
\mathcal{P}(A,N)$.}; finally we have }$U_{i}^{k}>U_{i}^{c}-\delta _{i}$ 
\textit{for all }$i$\textit{\ and all }$k\in \lbrack K]$\textit{.\smallskip }

\textbf{Proof} \textit{Statement }$i)$ Fix $P^{k}$ and write its congestion
profile simply as $(s_{x})_{x\in A}$, then fix an agent $i$ and her
allocation $(a,s_{a})$ at $P^{k}$; finally $[\sigma ^{c}]$ is the support of
the competitive congestion, containing $x$ iff $\sigma _{x}^{c}>0$.

Suppose first $s_{a}\geq 2$. By (\ref{14}) we have $s_{a}-1\leq \lfloor
\sigma _{a}^{c}\rfloor $ therefore $\sigma _{a}^{c}\geq 1$. We apply
repeatedly the monotonicity of $u_{i}(x,s)$ in $s$ and property (\ref{14}).%
\newline
For all $x\in \lbrack \sigma ^{c}]$: $u_{i}(a,s_{a}-1)\geq u_{i}(a,\lfloor
\sigma _{a}^{c}\rfloor )\geq u_{i}(a,\sigma _{a}^{c})\geq u_{i}(x,\sigma
_{x}^{c}\vee 1)\geq u_{i}(x,\lceil \sigma _{x}^{c}\rceil )$, where the last
inequality follows from $\sigma _{x}^{c}>0$. These inequalities imply that
agent $i$ prefers to $(a,s_{a}-1)$: at most $\lfloor \sigma _{a}^{c}\rfloor
-1$ less congested allocations at post $a$; at most $\lceil \sigma
_{x}^{c}\rceil -1\leq \lfloor \sigma _{x}^{c}\rfloor $ allocations at post $%
x $ if $\sigma _{x}^{c}>0$; and if $\sigma _{y}^{c}=0$ (\ref{14}) gives $%
u_{i}(a,s_{a})\geq u_{i}(y,1)$ so no allocation at $y$ improves $(a,s_{a})$
or $(a,s_{a}-1)$ for our agent $i$. We see that the number of allocations
improving $(a,s_{a}-1)$ is at most $(\sum_{x\in A}\lfloor \sigma
_{x}^{c}\rfloor )-1\leq \sum_{x\in A}\sigma _{x}^{c}-1=n-1$. As in
Proposition 1 we conclude that $(a,s_{a}-1)$ is top-$n$-fair, and we are
done.

In the case $s_{a}=1$ the argument is simpler. The allocation $(a,1)$ is the
best at post $a$; at any other post $x$ the inequality $u_{i}(a,1)\geq
u_{i}(x,\sigma _{x}^{c}\vee 1)$ implies it can be improved by at most $%
\lfloor \sigma _{x}^{c}\rfloor -1$ less congested allocations (or zero if $%
\sigma _{x}^{c}<1$). Then the allocation $(a,s_{a})$ itself is top-$n$%
-fair.\smallskip

\noindent \textit{Statement }$ii)$ Fix $P^{k},P^{\ell },a,b$ and $i\in
S_{a}^{k}\cap S_{b}^{\ell }$. By (\ref{14}) $a,b$ are both in $%
D(u_{i},\sigma ^{c})$ hence $u_{i}(a,\sigma _{a}^{c}\vee 1)=u_{i}(b,\sigma
_{b}^{c}\vee 1)$. By (\ref{14}) again and the definition of $\delta _{i}$ we
have $|u_{i}(a,s_{a}^{k})-u_{i}(a,\sigma _{a}^{c}\vee 1)|\leq \delta _{i}$
and $|u_{i}(b,s_{b}^{\ell })-u_{i}(b,\sigma _{b}^{c}\vee 1)|\leq \delta _{i}$%
, so the desired inequality follows.\smallskip

\noindent \textit{Statement }$iii)$ We fix $i\in S_{a}^{k}$ and $x$ then
combine three inequalities: $a\in D(u_{i},\sigma ^{c})$ gives $%
u_{i}(a,\sigma _{a}^{c}\vee 1)\geq u_{i}(x,\sigma _{x}^{c}\vee 1)$; next $%
|u_{i}(a,s_{a}^{k})-u_{i}(a,\sigma _{a}^{c}\vee 1)|\leq \delta _{i}$ follows
from $|s_{a}^{k}-\sigma _{a}^{c}|<1$ ((\ref{14})) and the definition of $%
\delta _{i}$; similarly $|u_{i}(x,s_{x}^{k})-u_{i}(a,\sigma _{x}^{c}\vee
1)|\leq \delta _{i}$.\smallskip

\noindent \textit{Statement }$iv)$ For the first part of the statement we
fix $P^{k}$ and $Q=(T_{x})_{x\in A}$ in $\mathcal{P}(A,N)$. There is at
least one post $b$ in the support of $Q$ ($T_{b}\neq \varnothing $) such
that $s_{b}^{k}\leq t_{b}$: otherwise $t_{b}<s_{b}^{k}$ for each $b$ s. t. $%
t_{b}\geq 1$, which contradicts $\sum_{A}t_{b}=\sum_{A}s_{b}^{k}=n$. Pick
such a post $b$, some $i\in T_{b}$, and let $a$ be the post assigned to $i$
by $P^{k}$ ($a=b$ is possible). By statement $iii)$ we have $%
u_{i}(a,s_{a}^{k})\geq u_{i}(b,s_{b}^{k}\vee 1)-2\delta _{i}$ and $%
u_{i}(b,s_{b}^{k}\vee 1)\geq u_{i}(b,t_{b})$ by our choice of $b$. So $Q$
does not improve $P^{k}$ by more than $2\delta _{i}$ for all agents in $%
T_{b} $.

For the second part we have $\sigma _{a}^{c}\geq 1$ for all $a$. Assume that
the utility profile of a deterministic assignment $Q=(T_{x})_{x\in A}$
Pareto dominates $(U_{i}^{c})_{i\in N}$. For any post $a$ in the support of $%
Q$ and agent $i\in T_{a}$, the definition of the competitive demand ((\ref%
{17})) and the choice of $Q$ imply $u_{i}(a,\sigma _{a}^{c})\leq
U_{i}^{c}\leq u_{i}(a,t_{a})$ and in turn $t_{a}\leq \sigma _{a}^{c}$ for
all $a$. As $\sum_{a}t_{a}=\sum_{a}\sigma _{a}^{c}$ we conclude $t=\sigma
^{c}$ and the inequalities above are equalities, so $Q$'s utility profile is
exactly $(U_{i}^{c})_{i\in N}$, and $Q$ does not dominate $(U_{i}^{c})_{i\in
N}$ after all.

Next for any $P^{k}$, any $i$ and $a$ such that $i\in S_{a}^{k}$, Definition
4 implies $s_{a}^{k}\leq \lceil \sigma _{a}^{c}\rceil <\sigma _{a}^{c}+1$
hence $u_{i}(a,s_{a}^{k})>u_{i}(a,\sigma _{a}^{c}+1)\geq u_{i}(a,\sigma
_{a}^{c})-\delta _{i}=U_{i}^{c}-\delta _{i}$ as desired. $\blacksquare
\smallskip $

The simplest examples of a fractional competitive assignments have only two
posts.

\textbf{Example 6 }\textit{Two posts and eight agents of three types, }$%
m=2,n=8$. The prefixes allow a single top-$n$-fair assignment%
\begin{equation*}
\begin{array}{ccc}
& a & b \\ 
\alpha \alpha \alpha \alpha & 8 & 0 \\ 
\beta \beta & 4 & 4 \\ 
\gamma \gamma & 0 & 8%
\end{array}%
\text{ \ \ top-}8\text{-fair }P_{1}:%
\begin{array}{cc}
a & b \\ 
\alpha \alpha \alpha \alpha & \beta \beta \gamma \gamma%
\end{array}%
\end{equation*}

$P_{1}$ is competitive if and only if\textit{\ }both $\beta $-s prefer (at
least weakly $(b,4)$ to $(a,4)$. We assume instead $(a,4)\succ _{\beta
}(b,4) $ for both $\beta $-s, so at $P_{1}$ they both envy post $a$. Their
cardinal utilities differ slightly: $\beta _{1}$ suffers comparatively less
than $\beta _{2}$ at $(a,5)$%
\begin{equation*}
\begin{array}{ccccc}
& (a,5) & (b,4) & (a,4) & (b,3) \\ 
u_{\beta _{1}} & 0 & 0 & 1 & 2 \\ 
u_{\beta _{2}} & -1 & 0 & 1 & 2%
\end{array}%
\end{equation*}

The competitive fractional congestion must load post $a$ more than post $b$: 
$\sigma ^{c}=(4+x,4-x)$ for some $x\geq 0$, therefore some deterministic
assignment implementing $\sigma ^{c}$ will violate top-$8$-fairness. By
statement $i)$ in Theorem 1 we have $x\leq 1$, so the demand of the $\alpha $%
-s and $\gamma $-s will not change. To meet property (\ref{13}) we need $%
D(u_{\beta _{i}},\sigma )=\{a,b\}$ for $i=1$ or $2$: the correct choice is $%
D(u_{\beta _{2}},\sigma )=\{a,b\}$, implying $x=\frac{1}{4}$ then $%
D(u_{\beta _{1}},\sigma )=\{a\}$.\footnote{%
Choosing $D(u_{\beta _{1}},\sigma )=\{a,b\}$ forces $x^{\prime }=\frac{1}{3}$
then $D(u_{\beta _{2}},\sigma )=\{b\}$, and (\ref{13}) fails.}

The lottery $\frac{3}{4}P_{1}+\frac{1}{4}P_{2}$ where%
\begin{equation*}
P_{2}=%
\begin{array}{cc}
a & b \\ 
\alpha \alpha \alpha \alpha \beta _{2} & \beta _{1}\gamma \gamma%
\end{array}%
\end{equation*}%
implements $\sigma ^{c}=(4\frac{1}{4},3\frac{3}{4})$ as specified by Theorem
1.

Note that for agent $\beta _{2}$, conditional on being at post $a$ the
expected congestion there is $5$, not $4\frac{1}{4}$: reasoning
competitively $\beta _{2}$ accepts both posts at $\sigma ^{c}$ without
taking into account the impact of their own assignment to $a$ on the
congestion there. By property (\ref{14}) this discrepancy is at most one
unit of congestion, irrespective of the size of $n$ and $m$.

Comparing the competitive lottery $\frac{3}{4}P_{1}+\frac{1}{4}P_{2}$ and
the top-$n$-fair assignment $P_{1}$ (also the only FM equ. outcome), agent $%
\beta _{1}$ and the $\gamma $-s prefer the lottery while $\beta _{2}$ and
the $\alpha $-s have the opposite preference.\smallskip

\textbf{Example 3 continued from section 3}. $m=3,n=12$.

The two types of agents have the following $12$-prefixes%
\begin{equation*}
\begin{array}{cccc}
& a & b & c \\ 
\alpha \alpha \alpha \alpha \alpha \alpha & 6 & 4 & 2 \\ 
\beta \beta \beta \beta \beta \beta & 2 & 4 & 6%
\end{array}%
\end{equation*}%
Suppose ordinal preferences are captured by the slack utilities (\ref{16}).
Their vNM extension to real valued congestion (\ref{24}) is still linear in
congestion: $u_{i}(a,\sigma _{a})=\lambda _{ia}-\sigma _{a}$.

We checked in section 3 that no deterministic assignment is competitive. The
unique competitive fractional congestion respects the symmetries of the
problem (this is a general property following its definition by the fixed
point property (\ref{13})), so we look for $\sigma ^{c}=(x,y,x)$ with $%
2x+y=12$. The only choice generating the demands $D(u_{\alpha },\sigma
)=\{a,b\}$, $D(u_{\beta },\sigma )=\{b,c\}$ is $\sigma ^{c}=(4\frac{2}{3},2%
\frac{2}{3},4\frac{2}{3})$.

To implement $\sigma ^{c}$ we must combine top-$12$-fair deterministic
assignments where $s_{a},s_{c}\in \{4,5\}$ and $s_{b}\in \{2,3\}$, which
leaves exactly three choices 
\begin{equation*}
\begin{array}{cccc}
& a & b & c \\ 
P_{2} & \alpha \alpha \alpha \alpha \alpha & \alpha \beta & \beta \beta
\beta \beta \beta \\ 
P_{4} & \alpha \alpha \alpha \alpha \alpha & \alpha \beta \beta & \beta
\beta \beta \beta \\ 
P_{5} & \alpha \alpha \alpha \alpha & \alpha \alpha \beta & \beta \beta
\beta \beta \beta%
\end{array}%
\end{equation*}%
Only $P_{2}$ is symmetric, it was already introduced in (\ref{43}).

The lottery $L^{c}=\frac{1}{3}P_{2}+\frac{1}{3}P_{4}+\frac{1}{3}P_{5}$ (plus
a uniform mixing of the $\alpha $-s and of the $\beta $-s in their
respective roles) is in this case the unique implementation of $\sigma ^{c}$.

The expected total utility of the $\alpha $-s is $\frac{1}{3}(7+6+10)$ so
each agent's expects the utility $1.28$. But the symmetric and top-$12$-fair
assignment%
\begin{equation*}
P_{3}:%
\begin{array}{ccc}
a & b & c \\ 
\alpha \alpha \alpha \alpha & \alpha \alpha \beta \beta & \beta \beta \beta
\beta%
\end{array}%
\end{equation*}
collects more expected utility. After the uniform mixing inside each type,
everyone gets $1.33$.

The trade-off is between gathering more surplus at $P_{3}$ but allowing $%
s_{b}=4$ so that the agents assigned to $b$ are (ex ante and ex post)
envious by a margin of $2$ units of congestion, versus losing some surplus
at the competitive lottery $L^{c}$ while generating no ex ante envy and ex
post envy only at the level of $1$ unit of congestion.

\textit{Remark}{\large \ }\textit{In the non congested assignment model
another approach to randomisation extends each deterministic ordinal
preference ordering to its (incomplete) stochastic dominance ordering of
random allocations; More popular in this century; than cardinal vNM.}

\textit{In the congested model this ordinal definition of competitiveness
fails because it is vastly underspecified: if we fix an arbitrary congestion
profile }$(s_{x})$\textit{\ we can force the PS algorithm (\cite{BM}, \cite%
{KS}, \cite{Bog}) to implement it\footnote{%
For each post $x$ we construct $s_{x}$ copies of the allocation $(x,s_{x})$
and assign (randomly) these $n$ objects to the $n$ agents. } and easily
select an envy free and ordinally efficient random assignment, but it is not
clear what deeper principle should guide the choice of that congestion
profile.}

\section{concluding comments}

We submit that our competitive analysis applies to a rich family of
congested assignment problems and delivers an efficient solution built upon
compelling fairness principles.

\paragraph{main results}

Whether congestion is anonymous or weighted, the canonical guarantee
eliminates for each participant all but the best $\frac{1}{m}$-th quantile
of the feasible allocations, where $m$ is the number of posts (Propositions
1 and 2).

Competitiveness, when it exists in the deterministic version of either
model, identifies an essentially single assignment in terms of congestion
and welfare. Combining efficiency with natural ex ante and ex post fairness
properties, it is then a compelling normative solution to the congested
assignment problem (Propositions 3 and 4).

After elicitating vNM cardinal utilities rather than simpler ordinal
preferences, the randomised competitive demand is unique (Lemmas 1 and 3)
and implemented by a lottery over competitive deterministic assignments
(Lemmas 2 and 4). In the anonymous case the latter are close to each other,
as well as approximately top-$n$-fair, competitive, and efficient (Theorem
1).

\paragraph{two open questions}

1). The addition of post-specific lower and upper bounds on congestion is
natural in many of the motivating examples discussed in the Introduction.
Under anonymous congestion we can easily generalise the concept of top-$n$%
-fair guarantee but its interpretation is a bit more involved. For agent $i$
the profile $\lambda _{i}$ must satisfy (\ref{1}) as well as the bounds: $%
\gamma _{a}^{-}\leq \lambda _{ia}\leq \gamma _{a}^{+}$ for all $a$. So the
number of allocations that may or may not be accepted by $\lambda _{i}$ is $%
\sum_{a}(\gamma _{a}^{+}-\gamma _{a}^{-})$ and $\lambda _{i}$ captures the
best $(n-\sum_{a}\gamma _{a}^{-})$ of these. Then Proposition 1 goes through
exactly as before.

Competitiveness is much harder to adapt. Its definition generalises when we
have only upper bounds, but we lose the critical fixed point argument in
Lemma 1 for the existence of a randomised competitive congestion profile.

With lower bounds it is not clear what the definition of competitiveness
should be: how to account for the situation where a lower bound $\gamma
_{a}^{-}$ forces the agents to populate a post $a$ that they unanimously
loath?\smallskip

2). A natural \textquotedblleft dual\textquotedblright\ domain of
preferences views congestion as strictly desirable: $(a,s+1)\succ (a,s)$ for
all $a$ and $s$. The top-fairness idea still works (at least under anonymous
congestion) but with much less bite. Interpret the report $\lambda _{i}$
satisfying (\ref{1}) as \textquotedblleft agent $i$ accepts $(a,s)$ only if $%
s\geq \lambda _{ia}$\textquotedblright .These constraints over all agents
are jointly feasible and allow each agent to veto at most $n-1$ allocations,
avoiding only the \textit{lowest} $\frac{1}{m}$ quantile of their
preferences. The proof mimicks that of Proposition 1 by switching a couple
of signs.

The Definition 3 of a competitive assignment is unchanged but no longer
identifies a unique solution for deterministic assignments (e. g. when
everyone cares only about joining the most congested post). Proposing and
justifying in the competitive spirit a fair randomised compromise in the
\textquotedblleft good congestion\textquotedblright\ model is a challenging
question

\section{appendix}

\subsection{proof of Proposition 2}

\textit{After proving that in any problem }$(A,N,\succeq ,w)$\textit{\ there
exists at least one top-}$\frac{1}{m}$\textit{-fair\ assignment }$P$\textit{%
, we state and prove the maximality property of the canonical
guarantee.\smallskip }

\textbf{Existence}\textit{. }As in Proposition 1 we combine a greedy
algorithm and an induction on the number of agents. We can clearly get rid
of the \textquotedblleft inactive\textquotedblright\ posts s. t. $\lambda
_{ia}=0$ for all $i$, so we still write $A$ for the set of active posts. We
pick any post $a$ and construct a set $S\subseteq A$ s. t.%
\begin{equation}
\forall i\in S:\lambda _{ia}\geq w_{S}\text{ and }\forall j\in N\diagdown
S:\lambda _{ja}<w_{S}+w_{j}  \label{15}
\end{equation}

Label the agents from $1$ to $n$ so that $\lambda _{ia}\geq \lambda
_{(i+1)a} $ for all $i\in \lbrack n-1]$ where $\lambda _{1a}>0$ because $a$
is active. For any two disjoints subsets $S,T$ in $A$ we say that $S$
rejects $T$ if $\lambda _{ia}<w_{S}+w_{T}$ for some $i\in S$; otherwise we
say that $S$ accepts $T$. Note that if all labels in $S$ are (weakly)
smaller than all in $T$ and $S$ rejects $T$, then $T$ rejects $S$ as well;
and $S$ accepts $T$ if $T$ accepts $S$. We construct $S$ recursively: in
each step we either find $S$ or add one agent to the provisional set.

\textbf{step }$1$. If for all $j\geq 2$ the set $\{j\}$ rejects $\{1\}$ then 
$S=\{1\}$ meets (\ref{15}) and we are done. Otherwise we pick the smallest
label $\ell ^{1}\geq 2$ accepting $\{1\}$, which implies that $\{1\}$
accepts $\{\ell ^{1}\}$ as well, and we form the provisional set $%
S^{1}=\{1,\ell ^{1}\}$. If $\ell ^{1}=n$ we are done by choosing $S^{1}$ so
going into step 2 we have $\ell ^{1}<n$.

\textbf{step} $k+1$. The subset $S$ has not yet been found therefore $\ell
^{k}$, the latest pick in $S^{k}$, is smaller than $n$. By construction $%
\lambda _{\ell ^{k}a}\geq w_{S^{k-1}}+w_{\ell ^{k}}=w_{S^{k}}$ so $\lambda
_{ia}\geq w_{S^{k}}$ for all $i\in S^{k}$. Moreover all agents $j<\ell ^{k}$
outside $S^{k}$ have rejected some earlier $S^{k^{\prime }}$, so they also
reject the larger set $S^{k}$.

If all agents $j>\ell ^{k}$ reject $S^{k}$ as well we are done by choosing $%
S^{k}$. Otherwise we pick the smallest label $\ell ^{k+1}$ after $\ell ^{k}$
s.t. $\{\ell ^{k+1}\}$ accepts $S^{k}$: this implies that $S^{k}$ accepts $%
\{\ell ^{k+1}\}$ as well (recall $\lambda _{\ell ^{k}a}\geq \lambda _{\ell
^{k+1}a}$) so we set $S^{k+1}=S^{k}\cup \{\ell ^{k+1}\}$ and we have $%
\lambda _{ia}\geq w_{S^{k+1}}$ for all $i\in S^{k+1}$. We are done if $\ell
^{k+1}=n$ otherwise we go to the next step. When this process stops we have
found $\widetilde{S}$ meeting (\ref{15}).\smallskip

We assign $\widetilde{S}$ to post $a$, and consider the residual problem in $%
\widetilde{A}=A\diagdown \{a\}$, $\widetilde{N}=N\diagdown \widetilde{S}$
with total congestion $\widetilde{W}=W-w_{S}$. For each $j\in \widetilde{A}$
such that $\lambda _{ja}>0$ inequality (\ref{15}) and equation (\ref{6})
imply%
\begin{equation*}
\lambda _{j\widetilde{A}}>\lambda _{jA}-(w_{S}+w_{j})=W-w_{S}+([[\lambda
_{j}]]-2)w_{j}=\widetilde{W}+([[\widetilde{\lambda }_{j}]]-1)w_{j}
\end{equation*}%
where in $\widetilde{\lambda }_{j}$ we drop $\lambda _{ja}$. For each $j\in 
\widetilde{A}$ such that $\lambda _{ja}=0$ inequality (\ref{5}) has no bite
and equation (\ref{6}) gives%
\begin{equation*}
\lambda _{j\widetilde{A}}=W+([[\lambda _{j}]]-1)w_{j}>\widetilde{W}%
+([[\lambda _{j}]]-1)w_{j}=\widetilde{W}+([[\widetilde{\lambda }%
_{j}]]-1)w_{j}
\end{equation*}

The induction argument shows that in the reduced problem there is a top-$%
\frac{1}{m-1}$-fair assignment of $N\diagdown S$ to $\widetilde{A}$%
.\smallskip

The \textbf{Maximality} property: \textit{Fix }$A,W$\textit{, an agent }$%
i^{\ast }$\textit{\ with weight }$w_{i^{\ast }}$\textit{\ and an arbitrary }$%
(W-w_{i^{\ast }})$\textit{-prefix} $\lambda _{i^{\ast }}$\textit{\ ((\ref{6}%
)). Then\ for any }$a$ s. t. $\lambda _{i^{\ast }a}>w_{i^{\ast }}$ \textit{%
there is a set of agents }$M$ \textit{not containing }$i^{\ast }$\textit{\
with weights }$w_{i}$ \textit{s. t.} $\sum_{i\in M}w_{i}=W-w_{i^{\ast }}$%
\textit{\ and a prefix }$\lambda _{i}$ \textit{((\ref{6}))} \textit{for each}
$i\in M$, \textit{s. t. in any top-}$n$\textit{-fair\ assignment of the }$%
M\cup \{i^{\ast }\}$\textit{\ problem agent }$i^{\ast }$\textit{\ is at a
post }$a$\textit{\ where the congestion is arbitrarily close to }$\lambda
_{i^{\ast }a}$.\smallskip

Proof: We\ fix $W$, $i^{\ast }$, the $(W-w_{i^{\ast }})$-prefix\textit{\ }$%
\lambda _{i^{\ast }}$ and $a$ as in the statement. The set $M$ contains one
agent $i_{b}$ for each post $b$ s. t. $\lambda _{i^{\ast }b}\geq w_{i^{\ast
}}$, in particular one $i_{a}$. The size of $M$ is $[[\lambda _{i^{\ast }}]]$%
. The weights are $w_{i_{b}}=\lambda _{i^{\ast }b}-w_{i^{\ast }}+\varepsilon 
$ if $b\in M\diagdown a$ and $w_{i_{a}}=\lambda _{i^{\ast }a}-w_{i^{\ast
}}-([[\lambda _{i^{\ast }}]]-1)\varepsilon $, where $\varepsilon >0$ is
small enough that $w_{i_{a}}>0$. Clearly $w_{i^{\ast }}+\sum_{b\in
M}w_{i_{b}}=W$. Suppose now that each $i_{b}$ (including $i_{a}$) is
single-minded on post $b$: $\lambda _{i_{b}b}=W$, $\lambda _{i_{b}x}=0$ for $%
x\neq b$. Consider a top-$\frac{1}{m}$-fair assignment of $M\cup \{i^{\ast
}\}$ to $A$: for each $i_{b}$ other than $i_{a}$ the inequality $%
w_{i_{b}}+w_{i^{\ast }}>\lambda _{i^{\ast }b}$ implies that $i_{b}$ is alone
at $b$, therefore $i^{\ast }$ share $a$ with $i_{a}$ where the congestion $%
\lambda _{i^{\ast }a}-([[\lambda _{i^{\ast }}]]-1)\varepsilon $ is
arbitrarily close to $\lambda _{i^{\ast }a}$. $\blacksquare $

\subsection{proof of Proposition 4}

It mimicks that of Proposition 3. Given the assignment $P=(S_{x})_{x\in A}$
for simplicity we write the congestion $w_{S_{x}}$ simplys as $s_{x}$. Now
agent $i$'s competitive demand at the congestion profile $s$ is\newline
$D(i;s)=\{a|(a,s_{a}\vee w_{i})\succeq _{i}(x,s_{x}\vee w_{i})$ for all $%
x\in A\}$, and $P$ is competitive iff $a\in D(i;s)$ whenever $i\in S_{a}$.

\textit{Statement }$i)$ \textit{Unique competitive congestion}. At first we
do not assume in $(A,N,\succeq ,w)$ either semi-strict preferences or a
crowded competitive assignment.

Fix $P=(S_{x})_{x\in A}$ and $P^{\ast }=(S_{x}^{\ast })_{x\in A}$ both
competitive and such that $s\neq s^{\ast }$. Partition $A$ as $B^{\ast }\cup
C\cup B$ where $B^{\ast }=\{a:s_{a}<s_{a}^{\ast }\}$ , $B=\{a:s_{a}^{\ast
}<s_{a}\}$ , $C=\{a:s_{a}^{\ast }=s_{a}\}$, and $B$, $B^{\ast }$ are both
non empty. We define in two equivalent ways the set $A^{\ast }=\{a\in
B^{\ast }:s_{a}^{\ast }>w_{i}$ for all $i\in S_{a}^{\ast }\}=\{a\in B^{\ast
}:|S_{a}^{\ast }|\geq 2\}$.

We claim first (as in the proof of Proposition 3) that if the demand $%
D(i;s^{\ast })$ of some agent $i$ intersects $A^{\ast }$, then $%
D(i;s)\subseteq A^{\ast }$. Suppose, to the contrary\newline
$a\in D(i;s^{\ast })\cap A^{\ast }$ while $b\in D(i;s)\cap A\diagdown
A^{\ast }$. By definition of the demands and of $B^{\ast }$ we have%
\begin{equation*}
(a,s_{a}^{\ast })\succeq _{i}(b,s_{b}^{\ast }\vee w_{i})\text{ and }%
(b,s_{b}\vee w_{i})\succeq _{i}(a,s_{a}\vee w_{i})\succ _{i}(a,s_{a}^{\ast })
\end{equation*}%
where the last relation is strict because $s_{a}\vee w_{i}<s_{a}^{\ast }$.
Note that we cannot replace $s_{b}\vee w_{i}$ by $s_{b}$ because $i$ may not
be in $S_{b}$.

If $b\notin B^{\ast }$ we have $s_{b}\vee w_{i}\geq s_{b}^{\ast }$ so we add 
$(b,s_{b}^{\ast }\vee w_{i})\succeq _{i}(b,s_{b}\vee w_{i})$ to the above
preferences to get a contradiction. If $b\in B^{\ast }$ then $|S_{b}^{\ast
}|=1$ because $b\notin A^{\ast }$, therefore $s_{b}<s_{b}^{\ast }=w_{i}$ so
that $(b,s_{b}\vee w_{i})=(b,s_{b}^{\ast }\vee w_{i})$ and we reach again a
contradiction. The claim is proved. As in the proof of Proposition 1 we
check next that $A^{\ast }$ must be empty: every $i\in {\Large \cup }_{a\in
A^{\ast }}S_{a}^{\ast }$ is s. t. $D(i;s)\subseteq A^{\ast }$ therefore $%
\sum_{a\in A^{\ast }}s_{a}\geq \sum_{a\in A^{\ast }}s_{a}^{\ast }$, which
contradicts $s_{a}^{\ast }>s_{a}$ in $B^{\ast }$. So for every $a\in B^{\ast
}$ we have $S_{a}^{\ast }=\{i\}$. Exchanging the roles of $P$ and $P^{\ast }$
we see that $S_{b}=\{i\}$ for all $b\in B$.

If at least one competitive assignment is crowded we take it as one of $%
P,P^{\ast }$: this contradicts the existence of $B$ and $B^{\ast }$ and we
are done. We continue the proof when we only know that preferences are
semi-strict. Writing $T^{\ast }={\Large \cup }_{a\in B^{\ast }}S_{a}^{\ast }$
and $T={\Large \cup }_{a\in B^{\ast }}S_{a}$ we have $w_{T^{\ast
}}=\sum_{a\in B^{\ast }}s_{a}^{\ast }>\sum_{a\in B^{\ast }}s_{a}=w_{T}$,
implying that $T^{\ast }\diagdown T$ is not empty: it contains some agent $%
i\in S_{a}^{\ast }\cap S_{b}$ where $a\in B^{\ast }$ and $b\in C\cup B$. By
definition of the demands and of the partition of $A$ we have%
\begin{equation*}
(a,w_{i})=(a,s_{a}^{\ast })\succeq _{i}(b,s_{b}^{\ast }\vee w_{i})\text{ and 
}(b,s_{b})\succeq _{i}(a,s_{a}\vee w_{i})=(a,w_{i})
\end{equation*}%
where the last equality is from $s_{a}<s_{a}^{\ast }=w_{i}$. If $b\in B$ we
have $w_{i}=s_{b}>s_{b}^{\ast }$ therefore $(b,s_{b}^{\ast }\vee
w_{i})\simeq _{i}(b,s_{b})$; this indifference still holds if $b\in C$
because in that case $s_{b}^{\ast }=s_{b}\geq w_{i}$: we conclude that all
preferences above are indifferences in particular $(a,w_{i})\simeq
_{i}(b,s_{b})$ which contradicts the semi-strictness of preferences.

\textit{Statement }$i)$ \textit{Unique competitive welfare}. Fix $%
P=(S_{x})_{x\in A}$ and $P^{\ast }=(S_{x}^{\ast })_{x\in A}$ both
competitive; we just proved $s=s^{\ast }$. An agent $i$ is in some $%
S_{a}^{\ast }\cap S_{b}$ (where $a=b$ is possible); by definition of
competitiveness and the inequalities $w_{i}\leq s_{a},s_{b}$ we have $%
(a,s_{a})\succeq _{i}(b,s_{b})$ and $(b,s_{b})\succeq _{i}(a,s_{a})$.

\textit{Statement }$ii)$ Assume $P$ is competitive and Pareto dominated by $%
Q $. Pick any agent $i\ $who is in $S_{a}$ at $P$ and in $T_{b}$ at $Q$ ($a=b
$ is possible). We have%
\begin{equation}
(b,t_{b})\succeq _{i}(a,s_{a})\succeq _{i}(b,t_{b}\vee w_{i})  \label{39}
\end{equation}%
therefore $t_{b}\vee w_{i}\geq t_{b}$. Suppose $s_{b}<t_{b}$: taking into
account $w_{i}\leq t_{b}$, the inequality $s_{b}\vee w_{i}\geq t_{b}$
implies $w_{i}=t_{b}>s_{b}$, and the two preferences in (\ref{39}) are
indifferences, in particular $(b,w_{i})\simeq _{i}(a,s_{a})$ a contradiction
of semi-strict preferences if $a\neq b$. If $a=b$ then (\ref{39}) gives $%
(b,t_{b})\succeq _{i}(b,s_{b})$ contradicting $s_{b}<t_{b}$.

We have shown $s_{b}\geq t_{b}$ everywhere on the support of $Q$, so these
are all equalities. By replacing $s_{b}$ by $t_{b}$ in (\ref{39}) we get $%
(b,t_{b})\simeq _{i}(a,s_{a})$ a final contradiction.

\textit{Statement }$iii)$ Fix a competitive assignment $P$, a post $a$ in
its support and an agent $i\in S_{a}$. We use the preferences $%
(a,s_{a})\succeq _{i}(x,w_{S_{x}}\vee w_{i})$ for all $x\neq a$ to evaluate
the size of the set of allocations agent $i$ may prefer to $(a,s_{a})$: at
post $a$ the size is $s_{a}-w_{i}$; at $x$ such that $w_{S_{x}}\geq w_{i}$
the size is $w_{S_{x}}-w_{i}$; and $0$ at $x$ such that $w_{S_{x}}\leq w_{i}$%
. So the total is at most $W-w_{i}$. $\blacksquare $

\subsection{fractional competitiveness under weighted congestion}

Switching to cardinal vNM utility functions we only partially emulate the
results of the previous section. Two of the key results are preserved: the
existence of a competitive fractional congestion profile and its qualified
uniqueness (Lemma 3 below); the implementation of the competitive congestion
by one or more lotteries over deterministic assignments (Lemma 4 below).
However we lose the rounding and approximation results of Lemma 2 and
Theorem 1.

The congestion profile $\sigma =(\sigma _{a})_{a\in A}$ is in the simplex $%
\Delta ^{%
%TCIMACRO{\U{211d} }%
%BeginExpansion
\mathbb{R}
%EndExpansion
}(A,W)$ and each agent $i$ has a cardinal vNM utility function $u_{i}(a,z)$
over $A\times \lbrack w_{i},W]$, continuous and strictly increasing in $z$.
As before if $\sigma _{a}<w_{i}$ the congestion price of post $a$ to agent $%
i $ is in fact $w_{i}$ so we do not define $u_{i}(a,z)$ in the interval $%
0\leq z<w_{i}$. Given a problem $(A,N,u,w)$ agent $i$'s competitive demand
at $\sigma $ is\newline
$D(u_{i},\sigma )=\arg \max_{x\in A}u_{i}(x,\sigma _{x}\vee w_{i})$ and the
fractional congestion profile $\sigma \in \Delta ^{%
%TCIMACRO{\U{211d} }%
%BeginExpansion
\mathbb{R}
%EndExpansion
}(A;W)$\ is competitive (Comp) iff $\sigma \in \sum_{i\in N}w_{i}\cdot
\Delta ^{%
%TCIMACRO{\U{211d} }%
%BeginExpansion
\mathbb{R}
%EndExpansion
}[D(u_{i},\sigma );1]$.

The (not necessarily unique) corresponding semi-stochastic matrix $\Pi $
meets (\ref{25}), but (\ref{23}) is replaced by%
\begin{equation}
\sigma _{a}=\sum_{i\in N}w_{i}\pi _{ia}\text{ for all }a\in A  \label{42}
\end{equation}

The same fixed point argument as in Lemma 1 proves that a solution of (\ref%
{42}) exists. To prove that it is unique we must strengthen the crowdedness
property of Definition 5 and prove a weaker uniqueness statement than in
Proposition 4 (subsection 5.2).\smallskip

\textbf{Definition 8 }\textit{The fractional congestion profile }$\sigma $%
\textit{\ is f-crowded in problem }$(A,N,u,w)$\textit{\ iff for all }$a$%
\textit{\ in its support (}$\sigma _{a}>0$\textit{) and each agent }$i$%
\textit{\ demanding }$a$\textit{\ (}$a\in D(u_{i},\sigma )$\textit{) we have 
}$\sigma _{a}>w_{i}$.\smallskip

\textbf{Lemma 3:}\textit{\ In any problem }$(A,N,\succeq ,w)$ \textit{there
is at most one f-crowded competitive congestion profile.}\smallskip

\textbf{Proof }\ Let $\sigma $,$\sigma ^{\ast }$ be two different f-crowded
competitive congestion profiles, and $B^{\ast }$ the non empty set of posts
s. t. $\sigma _{a}^{\ast }>\sigma _{a}$. We claim that if $D(u_{i},\sigma
^{\ast })$ contains a post $a\in B^{\ast }$ for some agent $i$ then $%
D(u_{i},\sigma )\subseteq B^{\ast }$. If, to the contrary, $D(u_{i},\sigma )$
contains $b$ outside $B^{\ast }$ we get the familiar inequalities $%
u_{i}(a,\sigma _{a}^{\ast })\geq u_{i}(b,\sigma _{b}^{\ast }\vee w_{i})$ and 
$u_{i}(b,\sigma _{b})\geq u_{i}(a,\sigma _{a}\vee w_{i})>u_{i}(a,\sigma
_{a}^{\ast })$ where the strict inequality is from $\sigma _{a}<\sigma
_{a}^{\ast }$ and the f-crowding assumption. If $b\notin B^{\ast }$ we have $%
\sigma _{b}^{\ast }\vee w_{i}\leq \sigma _{b}$ and derive a contradiction
from $u_{i}(b,\sigma _{b}^{\ast }\vee w_{i})\geq u_{i}(b,\sigma _{b})$.

Pick now some semi-stochastic matrices $\Pi $ and $\Pi ^{\ast }$ meeting (%
\ref{42}) for $\sigma $ and $\sigma ^{\ast }$ respectively, and use the
property $D(u_{i},\sigma ^{\ast })\cap B^{\ast }\neq \varnothing
\Longrightarrow D(u_{i},\sigma ^{\ast })\subseteq B^{\ast }$ to compute for
any $i\in N:\sum_{a\in A^{\ast }}\pi _{ia}^{\ast }>0\Longrightarrow
\sum_{a\in A^{\ast }}\pi _{ia}^{\ast }\leq \sum_{a\in A^{\ast }}\pi _{ia}$,
which implies $\sum_{a\in A^{\ast }}\sigma _{a}^{\ast }=\sum_{i\in
N}w_{i}(\sum_{a\in A^{\ast }}\pi _{ia}^{\ast })\leq \sum_{i\in
N}w_{i}(\sum_{a\in A^{\ast }}\pi _{ia}^{\ast })=\sum_{a\in A^{\ast }}\sigma
_{a}$, the final contradiction. $\blacksquare \smallskip $

The weighted congestion analog of Lemma 2 (section 6 for the anonymous
congestion case) is next.\smallskip

\textbf{Lemma 4 }\textit{In any problem }$(A,N,u,w)$\textit{\ we can
implement any competitive congestion profile }$\sigma ^{c}$ \textit{by one
or more list }$(\{P^{k}\}_{k=1}^{K},\mathcal{L})$ \textit{of }$K$\textit{\
deterministic assignments }$P^{k}$\textit{\ together with a lottery }$%
\mathcal{L}$ \textit{over} $[K]$\textit{, such that their expected
congestion is }$\sigma ^{c}$\textit{\ and they always assign the agents in
their competitive demands:}\newline
$i\in S_{a}^{k}\Longrightarrow a\in D(u_{i},\sigma ^{c})$.\smallskip

\textbf{Proof} Pick $\Pi $ realising the congestion $\sigma ^{c}$ in (\ref%
{42}) and apply the following version of the Birkhoff theorem: every
semi-stochastic matrix $\Pi $ is the convex combination of deterministic
matrices (all entries are $0$ or $1$) of which the support is contained in
that of $\Pi $. $\blacksquare \smallskip $

A final example illlustrates the limits of this result.

\textbf{Example 7: }\textit{Two posts, three agents}\textbf{\ }$m=2,n=3$%
\textit{\ and }$W=21$. The weights, $(W-w_{i})$-prefixes, and the two top-$%
\frac{1}{2}$-fair assignments are%
\begin{equation*}
\begin{array}{cccc}
& a & b & w \\ 
\alpha & 16 & 15 & 10 \\ 
\beta & 16 & 15 & 10 \\ 
\gamma & 9 & 13 & 1%
\end{array}%
\text{ \ \ \ }%
\begin{array}{ccc}
& a & b \\ 
P_{1} & \alpha & \beta \gamma \\ 
P_{2} & \beta & \alpha \gamma%
\end{array}%
\end{equation*}

If the preferences are represented by the slack utilities (\ref{16}) neither 
$P_{1}$ or $P_{2}$ is competitive because both $\alpha $ and $\beta $ prefer 
$(a,10)$ to $(b,11)$.

The fractional competitive congestion is $\sigma ^{c}=(11,10)$ because%
\newline
$D(\alpha ,\sigma ^{c})=D(\beta ,\sigma ^{c})=\{a,b\}$. It can be
implemented by two lotteries $\mathcal{L}_{1},\mathcal{L}_{2}$ combining $%
P_{1}$,$P_{2}$ with three not top-$\frac{1}{2}$-fair assignments:%
\begin{equation*}
\begin{array}{ccc}
& a & b \\ 
P_{3} & \alpha \gamma & \beta \\ 
P_{4} & \beta \gamma & \alpha \\ 
P_{5} & \alpha \beta & \gamma%
\end{array}%
\end{equation*}

Specifically: $\mathcal{L}_{1}=\frac{1}{4}P_{1}+\frac{1}{4}P_{^{2}}+\frac{1}{%
4}P_{3}+\frac{1}{4}P_{4}$, and $\mathcal{L}_{2}=\frac{9}{20}P_{^{1}}+\frac{9%
}{20}P_{^{2}}+\frac{1}{10}P_{5}$, . The former keeps congestion close to top-%
$\frac{1}{2}$-fair, but violates this property half of the time; the latter
only violates it with probability $\frac{1}{10}$ but then much more severely.

\end{document}